\documentclass[showkeys,aps,prr,amsmath,amssymb,reprint,floatfix,superscriptaddress]{revtex4-2}

\usepackage[caption=false]{subfig}
\usepackage[normalem]{ulem}
\usepackage{lipsum}
\usepackage[linkcolor=red, citecolor=blue, colorlinks=true, urlcolor=blue]{hyperref}
\usepackage{graphicx}
\usepackage{dcolumn}
\usepackage{bm}
\usepackage{xcolor}
\newcommand{\vj}{\boldsymbol{j}}
\newcommand{\vx}{\boldsymbol{x}}

\newcommand{\vk}{\boldsymbol{k}}

\newcommand{\vnabla}{\boldsymbol{\nabla}}
\newcommand{\vl}{\boldsymbol{l}}

\newcommand{\vrr}{\boldsymbol{r}}
\newcommand{\vp}{\boldsymbol{p}}

\graphicspath{{Figures/}}
\begin{document}

\title[]{Kardar-Parisi-Zhang universality in discrete two-dimensional driven-dissipative exciton polariton condensates}

\author{Konstantinos Deligiannis}
\affiliation{Univ. Grenoble Alpes, CNRS, LPMMC, Grenoble, 38000, France}
\author{Quentin Fontaine}
\affiliation{Universit\'e Paris-Saclay, CNRS, Centre de Nanosciences et de Nanotechnologies (C2N), 91120, Palaiseau, France}
\author{Davide Squizzato}
\affiliation{Univ. Grenoble Alpes, CNRS, LPMMC, Grenoble, 38000, France}
\affiliation{Dipartimento di Fisica, Universit\`a La Sapienza, Rome, 00185, Italy}
\affiliation{Istituto Sistemi Complessi, Consiglio Nazionale delle Ricerche, Universit\`a La Sapienza, Rome, 00185, Italy}
\author{Maxime Richard}
\affiliation{Univ. Grenoble Alpes, CNRS, Grenoble INP, Institut N\'eel, 38000 Grenoble, France}
\author{Sylvain Ravets}
\affiliation{Universit\'e Paris-Saclay, CNRS, Centre de Nanosciences et de Nanotechnologies (C2N), 91120, Palaiseau, France}
\author{Jacqueline Bloch}
\affiliation{Universit\'e Paris-Saclay, CNRS, Centre de Nanosciences et de Nanotechnologies (C2N), 91120, Palaiseau, France}
\author{Anna Minguzzi}
\affiliation{Univ. Grenoble Alpes, CNRS, LPMMC, Grenoble, 38000, France}
\author{L\'eonie Canet}
\affiliation{Univ. Grenoble Alpes, CNRS, LPMMC, Grenoble, 38000, France}
\affiliation{Institut Universitaire de France, Paris, 75000, France}

\begin{abstract}
The statistics of the fluctuations of quantum many-body systems are highly revealing of their nature.  In driven-dissipative systems displaying macroscopic quantum coherence, as exciton polariton condensates under incoherent pumping, the phase dynamics can be mapped to  the stochastic Kardar-Parisi-Zhang (KPZ) equation. However, in two dimensions (2D), it was theoretically argued that the KPZ regime may be hindered by the presence of vortices, and a non-equilibrium BKT behavior was reported close to condensation threshold. We demonstrate here that, when a discretized 2D polariton system is considered, universal KPZ  properties can emerge. We support our analysis by extensive numerical simulations of the discrete  stochastic generalized Gross-Pitaevskii equation. We show that the first-order correlation function of the condensate exhibits stretched exponential behaviors in space and time with critical exponents characteristic of the 2D KPZ universality class, and find that the related scaling function accurately matches the KPZ theoretical one, stemming  from functional Renormalization Group. We also obtain the distribution of the phase fluctuations and find that it is non-Gaussian, as expected for a KPZ stochastic process.

\end{abstract}
\maketitle

\section{Introduction}
\label{sec: introduction}

Scale invariance and universality are the most striking features unveiled by the study of equilibrium continuous phase transitions. They refer to the fact that in the vicinity of a critical point, systems of completely different nature can exhibit correlations in their order parameter fluctuations characterized by the same power-law, with the same universal critical exponents. 

The emergence of universality is well understood in systems at thermal equilibrium, but far less so out of equilibrium. In such systems, criticality can, for instance, emerge spontaneously during the time evolution of the system, a phenomenon known as self-organized criticality \cite{BakTangWiesenfeld1987}. A paradigmatic example of this feature is the mechanism of kinetic roughening of a stochastically growing interface \cite{HalpinHealyZhang1995, Krug1997}, for which the celebrated Kardar-Parisi-Zhang (KPZ) equation was originally derived \cite{KardarParisiZhang}. It was realized later on that a strikingly large variety of systems actually belong to the KPZ universality class \cite{Takeuchi2018}, such as switching fronts propagating in liquid crystals \cite{TakeuchiSano2012}, or the development of modern urban skylines under building constraints \cite{Najem2020} to cite a few.

Recently, KPZ universality was found in a surprising context:
in the phase dynamics of driven-dissipative exciton polariton (polariton) condensates. Polaritons are nonequilibrium quasi-particles that form within semiconductor microcavities operating in the strong coupling regime between bound electron-hole pairs (excitons) and cavity photons. Under non-resonant excitation, polaritons can spontaneously condense into a macroscopic coherent state \cite{Kasprzak2006, CarusottoCiuti2013,BlochCarusottoWouters2022}. In this state, the microscopic model describing the dynamics of the condensate phase has been shown to map onto the KPZ equation \cite{GladilinJiWouters2014, JiGladilinWouters2015, AltmanSiebererChenDiehlToner2015}. This has far reaching conceptual implications, as it shows that the nature of a polariton condensate is very different from its thermal equilibrium counterpart. KPZ universality was confirmed numerically in a one-dimensional (1D) polariton condensate, both in terms of scaling of spatiotemporal correlations \cite{HeSiebererAltmanDiehl2015}, and in terms of the phase fluctuation statistics \cite{SquizzatoCanetMinguzzi2018,Deligiannis2021}. In a polariton condensate, in contrast with more usual KPZ interfaces, space-time vortices can emerge stochastically due to the intrinsic $2\pi$ periodicity of the phase \cite{HeSiebererDiehl2017}. However, KPZ dynamics can co-exist with -- and remain robust to --  such defects. Indeed, KPZ scaling in a 1D polariton condensate has been reported experimentally \cite{Fontaine2021}, thus promoting polaritons as a compelling experimental platform to probe KPZ universality.

In two dimensions (2D), far fewer results have been obtained for the KPZ equation: no exact result is available, and mainly functional Renormalization Group calculations \cite{Canetetal2010,KlossCanetWschebor2012,KlossCanetWschebor2014,KlossCanetDelamotteWschebor2014} and  numerical simulations  \cite{ZabolitzkyStauffer1986, KerteszWolf1989, KimKosterlitz1989, ForrestTang1990, Kim1991, Tang1992, ChinNijs1999, KondevHenleySalinas2000, MarinariPagnaniParisi2000, AaraoReis2001, OdorLiedkeHeinig2009, KellingOdor2011, PagnaniParisi2015, MirandaAaraoReis2008, NewmanBray1996,HalpinHealy2012, OliveiraAlvesFerreira2013} have provided quantitative information. Experimentally, polaritons offer a unique opportunity to investigate KPZ physics, as they are well suited to form a two-dimensional condensate. The dynamics of its phase in 2D enables to study the 2D KPZ universality class releasing the stringent requirement of realizing a growing surface embedded in a three-dimensional physical space. In this work we address theoretically and numerically the emergence of KPZ physics in a non-resonantly driven discrete 2D polariton condensate. Our work is complementary to  the studies on coherent excitations ~\cite{Dagvadorj2015, ZamoraSiebererDunnettDiehlSzymanska2017, Dunnett2018}, the optical parametric oscillator \cite{FerrierEtAl2020}, and the quadratic driving scheme \cite{DiesselChiocchetta2022}, where the possibility of observing KPZ-like behavior was highlighted.  All of the above correspond to different models with respect to the one considered here. 

In 2D, equilibrium bosonic quantum fluids undergo the Berezinskii-Kosterlitz-Thouless (BKT) transition stemming from vortex-antivortex unbinding. Owing to the driven-dissipative nature of  polaritons, the behavior of vortices is  different: the attractive vortex-anti-vortex interaction is, for instance, dramatically damped at large distances, and can even change sign to become repulsive \cite{GladilinWouters2017, GladilinWouters2019A}, hindering their annihilation. In some specific non-equilibrium conditions, the vortex motion can also be self-accelerated, leading to the creation of additional vortex pairs \cite{GladilinWouters2017, GladilinWouters2019A}. The role of vortex unbinding in driven-dissipative condensates have also been studied using the duality with a theory for nonlinear electrodynamics coupled to charges   \cite{WachtelEtAl2016,SiebererWachtelAltmanDiehl2016}. Thus, vortex-filled phases are  also present in non-equilibrium systems and could be detrimental to the emergence of KPZ universality in 2D. This is supported by an argument based on perturbative Renormalization Group (RG), which suggests that the proliferation of vortices, which happens at spatial scale $L_v$, spoils KPZ stretched exponential decay of coherence characterized by a much longer length scale \cite{AltmanSiebererChenDiehlToner2015}. Moreover, several studies, based on both experiments and numerical simulations of 2D polariton condensates, have reported BKT-like types of behaviors, with specificities  that are related to the non-equilibrium regime
\cite{Szymanska2006,Caputo2017,GladilinWouters2019B,Comaron2021,GladilinWouters2021}.
 
An important feature of this system is that the dimensionality of the parameter space is large, such that a multitude of realistic regimes dominated by KPZ could still exist. For example,  the momentum-dependent loss rate of the polaritons plays an important role in the magnitude of the KPZ non-linearity $\lambda$, and hence in the possibility of reaching a KPZ regime. Indeed, KPZ scaling of spatiotemporal correlation functions were then found in some region of the parameter space, as well as vortex-free regimes \cite{MeiJiWouters2021}. Yet, this was obtained in the low noise limit, while a realistic noise input is typically larger and fixed by the polariton pump and losses.

In this work, we demonstrate the existence of a KPZ regime
for the phase dynamics of polariton condensates in
2D, using, for most parameters, values that can be realistically achieved in experiments.
We show that in this regime, the proliferation of topological defects such as spatial or spatiotemporal vortices can be limited by acting on the pump strength and by introducing a cut-off length scale via discretization of space thanks to the introduction of a lattice potential. Polariton lattices are routinely  created in semiconductor microcavities (see eg \cite{Schneider_2017} for a review).
We 
obtain the KPZ scaling in the spatiotemporal correlation function of the condensate wavefunction. This scaling is robust against some variations of the pump and of the lattice spacing. Furthermore, we compare the obtained scaling function with the one predicted for KPZ by functional Renormalization Group approaches, finding an excellent agreement. Finally, we provide compelling evidence for the 2D KPZ universality via the analysis of the fluctuations of the condensate phase. We compute the probability distribution function of the phase fluctuations, and show that it accurately matches the one obtained from large-scale numerical simulations of 2D KPZ systems \cite{HalpinHealy2012, HalpinHealy2013, OliveiraAlvesFerreira2013}.

The paper is organized as follows. The model studied, which is the generalized stochastic Gross-Pitaevskii equation for the dynamics of the condensate wavefunction, is introduced in Sec.~\ref{sec: model}. We recall the mapping from this equation to the KPZ equation in Sec.~\ref{sec: mapping}, and review the universal properties of the KPZ equation in 2D in Sec.~\ref{sec: 2DKPZ}. We describe the numerical simulations we performed, and we develop an analysis of the purely spatial and spatiotemporal vortices in Sec.~\ref{sec: simulations}. We present our results for the spatiotemporal correlation function in Sec.~\ref{sec: scaling}, and our results for the  distribution of the phase fluctuations in Sec.~\ref{sec: phase}. 

\section{Model for the polariton condensate dynamics} 
\label{sec: model}

The condensate wavefunction is described in 2D  by a classical field $\psi \equiv \psi(t, \vrr)$ at time $t$ and position $\vrr = (x,y)$. The condensate is populated through an incoherent excitonic reservoir of density $n_r(t, \vrr)$ which is continuously driven by an external pump at a rate $P$. The reservoir density decays into the condensate via stimulated relaxation at rate $R {\left\vert \psi \right\vert^2} $ and into other modes at rate $\gamma_r$. Under the assumption that the timescales for the reservoir and condensate dynamics are very different, the rate equation  for the reservoir can be integrated out, which leads to the generalized stochastic Gross-Pitaevskii equation for the condensate \cite{WoutersCarusotto2007b,WoutersSavona2009}
\begin{align}
i \hbar\partial_t \psi = &\Bigg[ - \left(\frac{\hbar^2}{2m} + V_\ell -i \dfrac{ \hbar\gamma_{l,2}}{2} \right)\nabla^2 + 2g_r \dfrac{P}{\gamma_r \left( 1 + {\left\vert \psi \right \vert^2}/{n_s} \right)}  \nonumber  \\
&   +i\frac{\hbar \gamma_{l,0}}{2} \left( \dfrac{p}{1 + {\left\vert \psi \right\vert^2}/{n_s}} -1 \right) + g \left\lvert \psi \right\rvert^2 \Bigg] \psi + \hbar \xi \,,
\label{eq: model}
\end{align}
where $m$ is the polariton effective mass, $V_\ell$ is the lattice potential, $g$  the polariton-polariton interaction strength, $g_r$ the repulsive interaction with excitons from the reservoir, $n_s = \gamma_r / R$, $p=PR/(\gamma_{l,0} \gamma_r)=P/P_{\textrm{th}}$, with $P_{\textrm{th}}$ the threshold pump rate for condensation. We consider a $k$-dependent loss rate of polaritons $\gamma_{l}(\vk) = \gamma_{l, 0} + k^2 \gamma_{l, 2}$, where $k$ is the polariton in-plane momentum \cite{Baboux2018,Fontaine2021}. Such a momentum dependence is equivalently introduced via a frequency-selective pump in the literature \cite{WoutersCarusotto2010, ChiochettaCarusotto2013, Comaron2021}. The noise $\xi$ is a complex variable with zero mean value, covariance $\left\langle \xi(t, \vrr) \xi^*(t^\prime, \vrr^\prime) \right\rangle = 2 \sigma \delta(\vrr-\vrr^\prime) \delta(t-t^\prime)$, and an intensity set by the drive and losses as $\sigma=\gamma_{l,0}(p+1)/4$.

\section{Mapping to the Kardar-Parisi-Zhang equation}
\label{sec: mapping}

As was shown in several studies, {\it e.g.} \cite{HeSiebererAltmanDiehl2015, SquizzatoCanetMinguzzi2018}, the dynamics of the phase of a spatially homogeneous driven-dissipative condensate can be mapped onto the KPZ equation. To show this, one expresses the condensate field in the density-phase representation $\psi(t, \vrr) = \sqrt{n(t, \vrr)}e^{i \theta(t, \vrr)}$.  Under the assumption that the density fluctuations $\delta n(t, \vrr) = n(t, \vrr)-\langle n\rangle$ are sufficiently smooth $\nabla \delta n \simeq 0$ and stationary $\partial_t \delta n \simeq 0$, one obtains for the phase dynamics the KPZ equation
\begin{equation}
\partial_t \theta = \nu \nabla^2 \theta + \dfrac{\lambda}{2} (\nabla \theta)^2 + \sqrt{D} \eta \,,
\label{eq: kpz}
\end{equation}
where the noise $\eta$ is real and has zero mean and covariance $\left \langle \eta(t, \vrr) \eta(t^\prime, \vrr^\prime) \right \rangle = 2 \delta(\vrr-\vrr^\prime) \delta(t-t^\prime)$. The KPZ parameters  are determined by the microscopic parameters of Eq.~(\ref{eq: model}) as
$\nu = {\gamma_{l, 2}}/{2} + a {\hbar}/{(2m)}$,
$\lambda = -{\hbar}/{m} + a {\gamma_{l,2}}$,
$D = {\sigma(1 + a^2 )}/{(2n_s(p-1))}$, where
$a = {2pgn_s}\left(1 - 2{Pg_r}/{(p^2 g \gamma_r n_s)} \right)/{\hbar\gamma_{l,0}} $.
Several remarks are required. 
First, the mapping is readily extended to the case of an external confinement see eg. \cite{Deligiannis2021}. 
Then, let us notice that for classical interfaces, the height field $h(\vrr,t)$ describes a $D$-dimensional manifold embedded in a $d=D+1$ space. The mapping hence relates the phase  $\theta(t, \vrr)$ of a condensate in $D=2$ spatial dimension to an interface of internal dimension two in a $d=2+1$ space. It holds in any dimension provided the previous assumptions are fulfilled. Second, another fundamental difference lies in the compactness of the phase, which is periodically defined $\theta\in(-\pi,\pi]$, whereas the height of an interface is an unbounded field $h\in(-\infty,\infty)$. This implies that the phase field can feature topological defects, such as phase jumps or vortices as already mentioned. While a compact version of the KPZ equation turns out to be relevant in some systems such as driven vortex lattices in disordered superconductors \cite{Aranson1998} or polar active smectic phases \cite{Chen2013}, we show that the behavior of the phase in the regime studied in this work belongs to the non-compact KPZ universality class, as it involves only a few sporadic vortices and the condensate phase can be uniquely unwound. 
As a third remark, let us mention  that several works introduce some anisotropy in the polariton effective mass $(m_\parallel,m_\perp)$, which leads through a similar mapping to the anisotropic extension of the KPZ equation \cite{AltmanSiebererChenDiehlToner2015}. However, as long  as $m_\parallel$ and $m_\perp$ have the same sign, the anisotropy is irrelevant in the Renormalization Group sense and the universal properties of the anisotropic KPZ dynamics are controlled by the isotropic fixed point. When they have opposite sign, the non-linearity $g_{\rm KPZ}=\lambda^2 D/\nu^3$ becomes irrelevant and the effective properties of the system are governed by the anisotropic Edwards-Wilkinson (Gaussian) fixed point \cite{Wolf1991, TauberFrey2002, KlossCanetWschebor2014}. Since we are interested in the nonlinear KPZ regime, we do not include any anisotropy, and consider an isotropic polariton dispersion relation.

\section{KPZ equation in 2D}
\label{sec: 2DKPZ}

While a wealth of exact results have been obtained for the KPZ equation in 1D, no exact results are available in 2D, except the exact exponent relation $\chi+z=2$ which stems from the statistical tilt symmetry of the KPZ equation and holds in all dimensions. The universal properties of the KPZ equation in 2D have been thoroughly studied using numerical simulations of discrete models belonging to the KPZ universality class, see {\it e.g.} Refs.~\cite{ZabolitzkyStauffer1986, KerteszWolf1989, KimKosterlitz1989, ForrestTang1990, Kim1991, Tang1992, ChinNijs1999, KondevHenleySalinas2000, MarinariPagnaniParisi2000, AaraoReis2001, OdorLiedkeHeinig2009, KellingOdor2011, PagnaniParisi2015}, or of the KPZ equation itself, {\it e.g.} Refs.~\cite{AmarFamily1990, GrossmanGuoGrant1991, MoserKerteszWolf1991, MirandaAaraoReis2008, NewmanBray1996}. The most precise estimates yield a roughness exponent $\chi\simeq 0.39$, $z=2-\chi \simeq 1.61$, and thus a growth exponent $\beta=\chi/z\simeq 0.24$ (we recall that in 1D, one has $\chi=1/2$ and $\beta=1/3$). The probability distribution of height fluctuations has also been computed numerically in Refs.~\cite{HalpinHealy2012,HalpinHealy2013,OliveiraAlvesFerreira2013}, which suggest the existence of different universality sub-classes, associated to different global geometries of the growth, as in 1D (see below).

On the analytical side, the KPZ  phase is described by a genuinely strong-coupling fixed-point in $D \geq 2$. Indeed, it was shown in \cite{Wiese1998} that the perturbative RG flow equation for the KPZ coupling $g_{\rm KPZ}$ can be resummed exactly, yielding an expression valid to all orders in perturbation theory in the vicinity of 2D. However, this expression fails to capture the strong-coupling KPZ fixed-point, yielding instead a run-away flow to infinity. Hence, let us emphasize that one cannot extract any quantitative information about the KPZ fixed-point in 2D or higher dimensions from the perturbative flow equation.

A non-perturbative RG calculation was devised using functional renormalisation group in Refs.~\cite{Canetetal2010,Canetetal2011,KlossCanetWschebor2012}, which allows accessing the strong-coupling KPZ fixed-point in all dimensions. It shows that the KPZ fixed-point is not connected to the Gaussian Edwards-Wilkinson fixed-point in any dimension, which explains the failure of perturbation theory, even resummed at all orders. The KPZ universal scaling function $F$ in 2D was calculated in \cite{KlossCanetWschebor2012} and will serve as our theoretical reference. It is defined as
\begin{equation}
C(\Delta t, \Delta \vrr) = C_0 {\Delta t}^{2\beta} F\left(y_0 \frac{|\Delta \vrr|}{\Delta t^{1/z}}\right)\,,
\label{eq: scaling}
\end{equation}
where $C$ is the two-point connected correlation function
\begin{align}
C(\Delta t, \Delta \vrr) =& \langle h(t+\Delta t, \vrr+\Delta \vrr) h(t, \vrr)\rangle \nonumber
\\ &-  \langle h(t+\Delta t, \vrr+\Delta \vrr)\rangle \langle h(t, \vrr)\rangle\, ,
\label{eq: defCh}
\end{align}
and $C_0$ and $y_0$ are non-universal normalisation constants. The scaling function has the following asymptotics,
\begin{equation}
F(y) \stackrel{y\to 0}{\longrightarrow}  F_0 \; , \quad  F(y) \stackrel{y\to \infty}{\sim}   F_\infty y^{2\chi}\,,
\label{eq: gy}
\end{equation}
where $F_0$ and $F_\infty$ are constants given in Appendix~\ref{app: Frond}.

\section{Numerical simulations and analysis of vortices} 
\label{sec: simulations}

\begin{figure}
\includegraphics[scale=1]{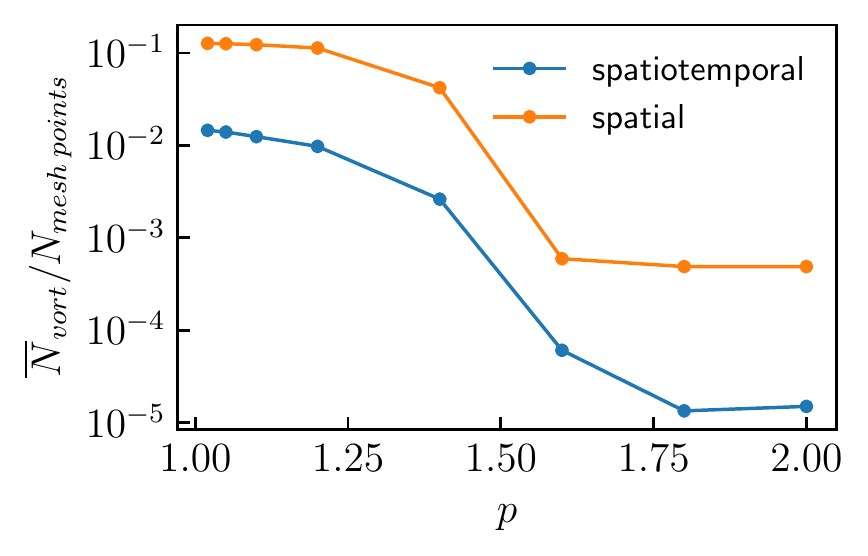}\\
\includegraphics[scale=0.875]{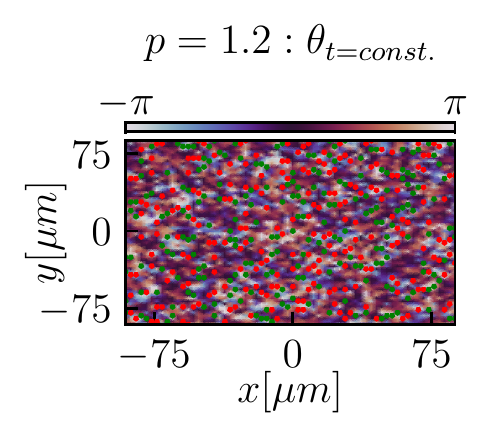}\hfill%
\includegraphics[scale=0.875]{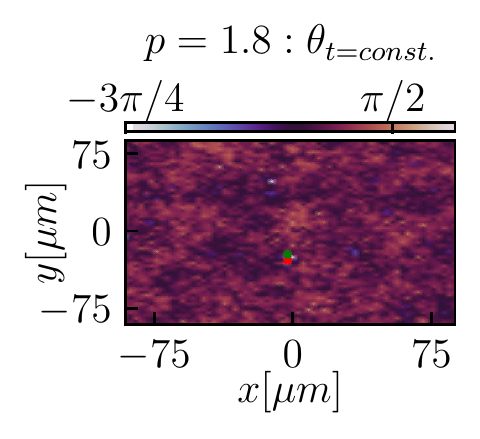}
\caption{Density of purely spatial and space-time vortices as a function of the pump power $p$, with typical phase configurations in the $(x,y)$ plane (vortex - green, anti-vortex - red) shown for $p=1.2$ (left) and $p=1.8$ (right) for a grid spacing of $dx=2.83 \mu m$.}
\label{fig: fig1}
\end{figure}
\begin{figure}
\includegraphics[scale=1]{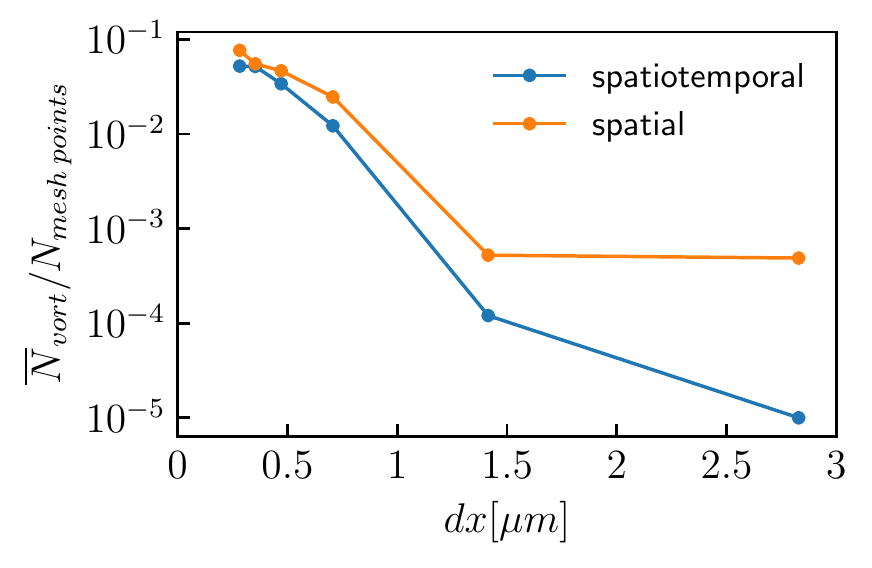}\\
\includegraphics[scale=0.865]{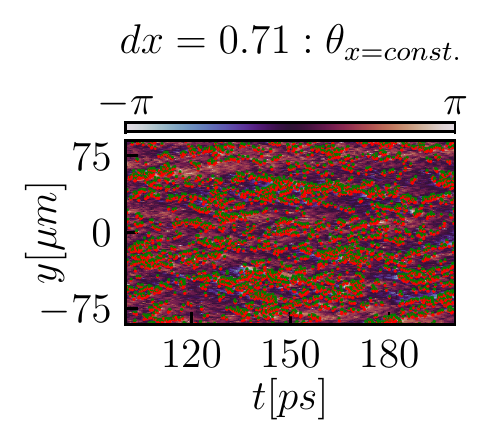}\hfill%
\includegraphics[scale=0.865]{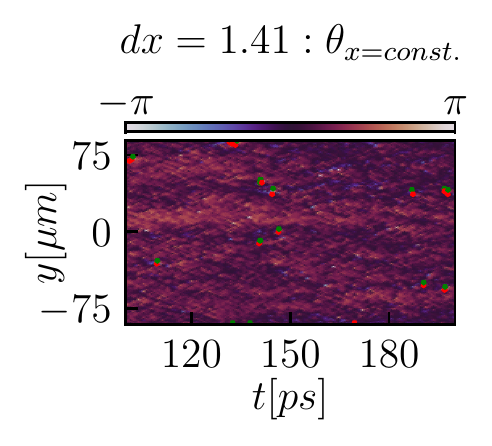}
\caption{Density of purely spatial and space-time vortices as a function of the lattice spacing, with typical phase configurations in the $(t,y)$ plane (vortex - green, anti-vortex - red) shown for a grid spacing of $dx=0.71\mu m$ (left) and $dx=1.41\mu m$ (right), for a fixed pump power $p=2$.}
\label{fig: fig2}
\end{figure}

In the deep lattice regime, we solve a discretized version of Eq.~(\ref{eq: model}) (see Appendix A for derivation and details)
\begin{align}
    i \hbar  \partial_t \psi_{\vj} =& - K ( \psi_{j_x+1,j_y} + \psi_{j_x-1,j_y} +\psi_{j_x,j_y+1} + \psi_{j_x,j_y-1} ) \nonumber \\ 
    &+ W_0  \psi_{\vj} + i \frac{{\gamma}_{l,0}}{2} \left(\frac{p}{1+|\psi_{\vj}|^2/{\cal N}_s}-1 \right) \psi_{\vj} + \hbar \xi_{\vj}
    \label{eq:discr-GPE}
\end{align}
We numerically integrate the gGPE~\eqref{eq:discr-GPE}
using a split-step procedure for each noise realization (see Appendix \ref{app: vortices} for details). We use a 2D $N\times N$ lattice grid
with spatial spacing $dx = 2.83 \mu m$, with $N=64$ for the analysis of the correlation functions, and $N=32, 64, 128$ for the study of the fluctuations distribution. 
In \cite{Fontaine2021},  where KPZ was observed in a 1D lattice,  the full dispersion of the system was approximated in the simulations by a parabolic dispersion, which provided a very accurate description. In 2D, we also approximate the dispersion of the lattice model
({\it e.g.} a cosine dispersion in the tight-binding regime)
 with a parabolic one. Indeed, the two dispersions are equivalent in the vicinity of $k=0$ and differ only at large $k$. Thus, we expect the difference between the two to occur mainly at small spatial scales, which lie below the ones relevant for KPZ dynamics, and which are not of direct interest in this study.
We choose the following parameters: effective mass $m^* = 8 \times 10^{-5} m_e$, where $m_e$ is the mass of the electron, corresponding to  Re$K/\hbar= 0.72 ps^{-1}$; $\gamma_{l, 0} = 0.31 ps^{-1}$ , $\gamma_r = \gamma_{l, 0} / 10$, $\gamma_{l, 2} = 0.1$ $\mu m^2 ps^{-1}$, yielding Im$K/\hbar= 0.05 ps^{-1}$. We note that the value of $\gamma_{l, 0}$ corresponds to a polariton lifetime $\tau = 3.2ps$, which is much too fast for significant equilibration to take place. 
Furthermore, we consider a non-interacting system with $g=0$, since this turned out to be a good description of the 1D experiment. We note that the presence of polariton-polariton ($g$) and polariton-reservoir ($g_r$) interactions are not fundamental for the existence of a KPZ regime, since the mapping still holds for $g=g_r=0$. In fact, the study reported in Ref.~\cite{MeiJiWouters2021} suggests that the development of the KPZ regime is favored by weak interactions. 
To  further reduce the parameter space to explore, we also set $g_r=0$, although we stress that in order to faithfully model the experimental conditions one should take a non-zero value for this term. The saturation density is fixed at ${\cal N}_s= 120$.
The numerical integration is performed with a time step of $dt = 0.16ps$ to study the scaling, and down to $dt = 0.0016ps$ to study the statistics of the fluctuations. This finer time discretization ensures that the phase can be uniquely unwrapped in time $\theta(t, \vrr) \in (-\pi, \pi] \rightarrow (-\infty, \infty)$. The phase is unwrapped at a fixed space-point by constraining the phase difference between two consecutive times to be less than $ 2\pi \zeta$. In practice, we choose  $\zeta$ not exactly one but $\zeta \lesssim 1$ due to the final time resolution. This procedure is robust as long as  $\zeta$ is typically greater than $0.5$.

In order to tentatively locate the KPZ regime, we first investigate the density of topological defects, which are either purely spatial vortices in the $(x,y)$ plane, or space-time vortices which have a non-vanishing projection of their vorticity in the $(t,x)$ and/or $(t,y)$ planes. The latter were shown to play an important role in 1D \cite{HeSiebererDiehl2017, Fontaine2021}. Maintaining the other parameters fixed, we vary the reduced pump $p$ from low $p\simeq 1$ (close to threshold) to moderately high $p=2$. For each value of $p$, we determine the density of purely spatial and space-time vortices in the stationary state, which is shown in Fig.~\ref{fig: fig1}, with typical phase configurations in the $(x,y)$ plane (similar maps are obtained in the $(t,x)$ and $(t,y)$ planes). The numerical procedure to detect the vortices and compute their density is detailed in Appendix~\ref{app: vortices}. We find that the number of vortices drastically decreases when increasing the pump power, and for $p\gtrsim 1.6$, very few spatial vortices are present, in agreement with Ref.~\cite{Comaron2021}, and also very few space-time vortices. Both appear only by pairs of closely located vortex and anti-vortex. We hence focus in the following on the highest value of the pump $p=2$, which appears as more favorable for observing KPZ dynamics. We show in Appendix~\ref{app: robust} that the KPZ regime is robust, to a certain extent, against decreasing the pump power.

The presence of topological defects turns out to be also sensitive to the discreteness of space, and we find that the presence of a lattice (of spacing $dx$) is favorable to observe the KPZ regime since the density of such defects is suppressed when increasing the grid  spacing $dx$, as shown in Fig.~\ref{fig: fig2}. For $dx \gtrsim 1\mu m$, the emergence of space-time vortices is rare and very short-lived. In the following, we choose $dx=2.83\mu m$, which is close to the spacing in experimental micro-pillar lattices \cite{Fontaine2021}. For this value, the fraction of both spatial and space-time vortices is below $10^{-3}$ and does not spoil the KPZ universal properties. Moreover, the core size of the spatial vortices was observed to be very narrow, of the order of the lattice spacing.

\section{KPZ scaling in spatiotemporal correlation functions}
\label{sec: scaling}

\begin{figure}[ht!]
\includegraphics[scale=1]{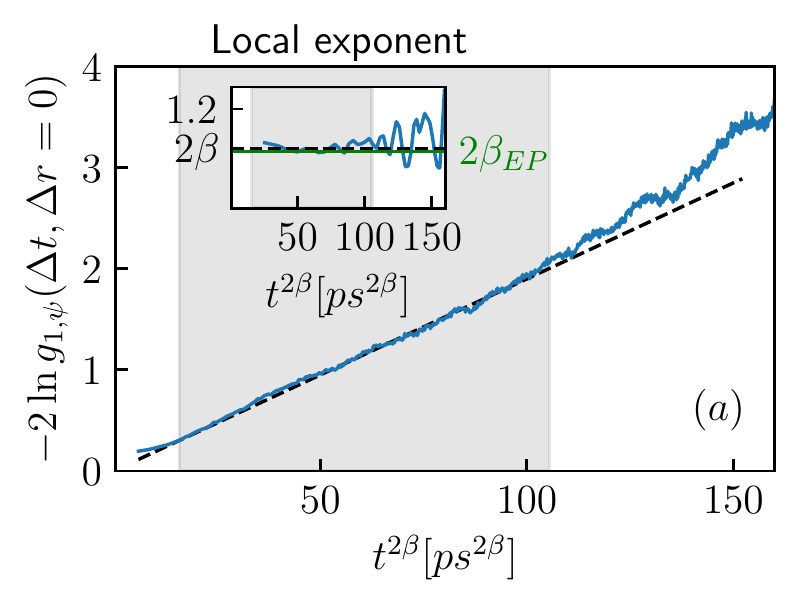}
\includegraphics[scale=1]{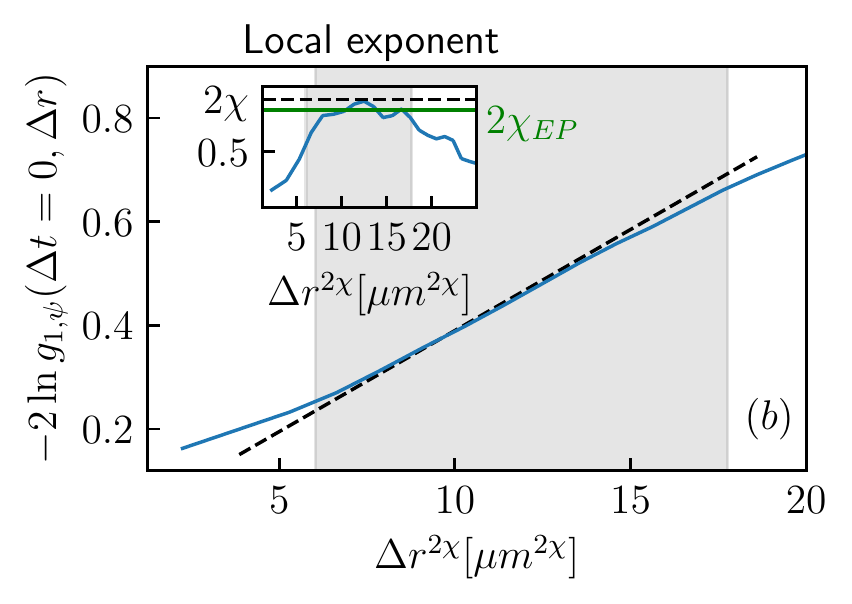}
\caption{(a) Equal-space and (b) equal-time correlation functions (blue plain lines) evaluated from $g_{1, \psi}$ as explained in the text, compared with the KPZ scaling laws  (black dashed lines). The numerical data follow the KPZ scaling for an extended range of time and space separations (indicated by the gray shade). The local exponents, which are computed from Eqs.~(\ref{eq: local_exponent1}, \ref{eq: local_exponent2}) are shown in the inset in both cases, together with their fits by a constant in the gray region (solid green lines, denoted $2\beta_{EP}$ and $2\chi_{EP}$) and the values from KPZ numerical simulations (black dashed line, denoted $2\beta$ and $2\chi$). Note that a running average has been used in order to smoothen the equal-space correlation function (i) before computing its derivative.}
\label{fig: scaling}
\end{figure}

Using our numerical simulations we calculate the first-order correlation function $g_{1, \psi}$ of the polariton condensate wavefunction, which is defined as
\begin{equation}
g_{1, \psi}(\Delta t, \Delta r) = \dfrac{ \left\langle \psi^*(t_0, \vrr_0) \psi(t_0 + \Delta t, \vrr_0 + \Delta \vrr) \right\rangle }{\sqrt{ \left\langle n(t_0, \vrr_0) \right\rangle} \sqrt{\left\langle n(t_0 + \Delta t, \vrr_0 + \Delta \vrr) \right\rangle}} \,,
\label{eq: g1}
\end{equation}
where $\left\langle \cdots \right\rangle$ denotes the ensemble average over noise realizations. Because of isotropy, $g_{1, \psi}$ only depends on the modulus $\Delta r = |\Delta \vrr|$, therefore we also average the data over circular shells of radius $\Delta r$ around $\vrr=0$.

Under the assumption that the density-phase and density-density correlations  are both negligible, the expression of $g_{1, \psi}$ becomes $g_{1, \psi}(\Delta t, \Delta r) \simeq \left\langle e^{i  \Delta\theta } \right\rangle$, where $\Delta\theta\equiv\theta(t_0 + \Delta t, \vrr_0 + \Delta \vrr) -\theta(t_0, \vrr_0)$.
Upon performing a cumulant expansion of $g_{1, \psi}$ to lowest order, one deduces that it is related to the connected correlation function $C$ of the phase as
\begin{equation}
-2 \ln \left[\lvert g_{1, \psi}(\Delta t, \Delta r)\rvert  \right]  = \big\langle \left[ \Delta\theta \right]^2 \big\rangle- \big\langle \Delta\theta\big \rangle^2 \equiv C(\Delta \vrr, \Delta t) \,.
\label{eq: cumulant_expansion}
\end{equation}
where $C$ is the analogue quantity as  Eq.~(\ref{eq: defCh}) for the interface. Note that $g_{1,\psi}$ constitutes a reliable observable to access the scaling behavior of the phase only provided the three previous assumptions are verified. This is not guaranteed a priori and has to be assessed. It was shown to be the case in 1D in the experimental conditions where the KPZ regime was evidenced~\cite{Fontaine2021}. They are also satisfied in our study, as shown in Appendix~\ref{app: robust}.

We first study the equal-time and equal-space correlation functions. If the phase follows the KPZ dynamics, then according to Eqs.~(\ref{eq: scaling}, \ref{eq: gy})
\begin{subequations}
\begin{align}
-2 \ln |g_{1, \psi}(\Delta t=0, \Delta r)| &\sim   \Delta r^{2\chi} \,,
\label{eq: scalinglaw1} \\
-2 \ln |g_{1, \psi}(\Delta t, \Delta r=0)| &\sim   \Delta t^{2\beta} \,,
\label{eq: scalinglaw2}
\end{align}
\end{subequations}
with $\chi \simeq 0.39$ and $\beta \simeq 0.24$. We observe the expected scaling behavior both in space and time, as shown in Fig.~\ref{fig: scaling}. For temporal correlations, the KPZ scaling extends for time differences spanning over a decade, from $3\times 10^2 - 1.5\times10^4 ps$, whereas for spatial correlations, the KPZ scaling is observed over a range from $10 - 40 \mu m$. The limited space range is related to the relatively small size of our system, with a condensate of approximately $90 \mu m$ radius. Note that the polariton interaction with the reservoir $g_r$ was here set to zero, but it is likely to play a role on the typical  space and time scales where KPZ dynamics dominates, as observed in 1D \cite{Fontaine2021}.

In order to provide a quantitative estimate of the critical exponents $\chi$ and $\beta$, we compute the local exponents, given by the following logarithmic derivatives  
\begin{subequations}
\begin{align}
\dfrac{d}{d \ln \Delta r}\Bigl[ \ln \Bigl(-2 \ln |g_{1, \psi}(\Delta t=0, \Delta r)| \Bigr) \Bigr]  &\sim {2\chi} \,,
\label{eq: local_exponent1} \\
\dfrac{d}{d \ln \Delta t}\Bigl[ \ln \Bigl(-2 \ln |g_{1, \psi}(\Delta t, \Delta r=0)| \Bigr) \Bigr] &\sim {2\beta} \, .
\label{eq: local_exponent2}
\end{align}
\end{subequations}
The result is shown in the insets of Fig.~\ref{fig: scaling}.
We fit the obtained local exponents by a constant in the appropriate spatial and temporal windows, which yields the values of the universal exponents $\beta_{EP} \simeq 0.22 \pm 0.06$ and $\chi_{EP} \simeq 0.36 \pm 0.04$. These values are in good agreement with the results from numerical simulations for the 2D KPZ universality class \cite{PagnaniParisi2015}.

\begin{figure}
\includegraphics[scale=1]{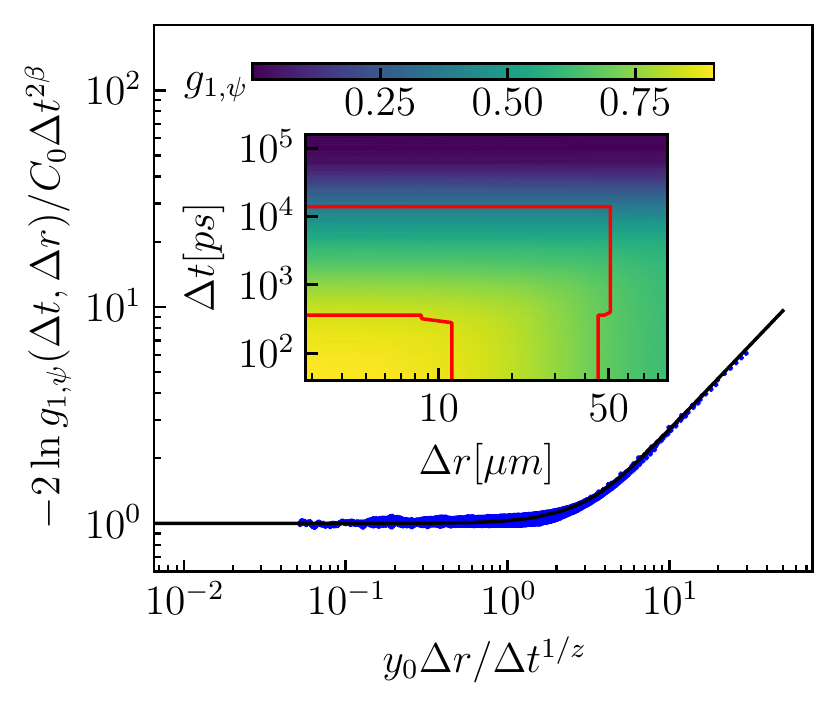}
\caption{Universal scaling function $F(y)$ from our numerical data (blue dots) and from the functional Renormalization Group calculation (black solid line). Inset: Space-time map of $g_{1, \psi}$, with the red contour delimiting the scaling region where data are selected to construct $F(y)$.}
\label{fig: gy}
\end{figure}

We now study the scaling properties of the spatiotemporal correlations over the whole relevant domain $(\Delta t, \Delta r)$. For this, we first select the data lying within the scaling regime, indicated in the inset of Fig.~\ref{fig: gy}. We then construct the scaling function $F(y)$ defined in Eq.~(\ref{eq: scaling}). The normalizations $C_0, y_0$ are determined from fitting the equal-time and equal-space correlation functions with the power-laws Eqs.~(\ref{eq: scalinglaw1}, \ref{eq: scalinglaw2}) in the appropriate ranges. 
Our results are shown in Fig.~\ref{fig: gy}, together with the theoretical scaling function calculated from functional Renormalization Group. We observe a collapse of the data onto a single curve, which matches the theoretical one with excellent accuracy. This result shows that the phase of the two-dimensional polariton condensate follows a KPZ effective dynamics over extended length and time scales.

\section{Statistical properties of phase fluctuations}
\label{sec: phase}

We next compute the probability distribution of the fluctuations of the unwrapped phase $\theta \equiv \theta(t, \vrr_0)$ at a fixed space point $\vrr_0$. In order to discuss the 2D results, we first briefly summarize what is known for a 1D classical interface  described by a height field $h$. In this case, the distribution of the reduced height fluctuations defined as $\tilde{h} = (h-v_\infty t)/(\Gamma t)^{1/3}$, with $v_\infty$ and $\Gamma$ non-universal parameters, is known exactly. In fact, the distribution of $\tilde h$ was shown to be universal but sensitive to the global geometry of the growth, which led to the definition of three main universality sub-classes. For flat, curved, or stationary geometries, the probability distribution is given by Tracy-Widom GOE (Gaussian Orthogonal Ensemble), Tracy-Widom GUE (Gaussian Unitary Ensemble) or Baik-Rains distributions, respectively \cite{PrahoferSpohn2000}. They all share the common feature of being non-Gaussian, with finite skewness and excess kurtosis. Strikingly, the three universality sub-classes can also be realized in the polariton condensate in 1D, where the role of global geometry can be emulated by appropriate external potentials \cite{Deligiannis2021}.

\begin{figure}[ht!]
\includegraphics[scale=1]{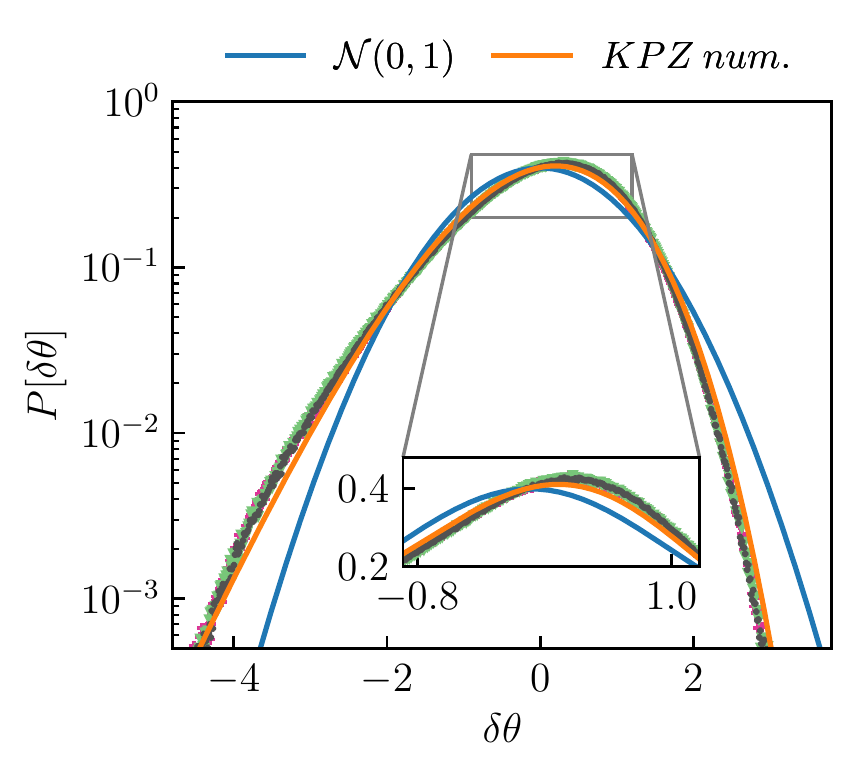}
\caption{Histogram of the centered and rescaled to unit-variance phase fluctuations $\delta \theta$ in the appropriate KPZ time window for different system sizes $L=90.5 \mu m, 181 \mu m, 362 \mu m$ (fuchsia, light green, dark gray symbols, respectively). 
Each curve comprises of approximately $8\times10^6$ data for each system size. The centered and unit-variance Gaussian distribution (denoted $\mathcal{N}(0,1)$) is also shown for comparison, together with the distribution found from KPZ large-scale simulations (denoted KPZ num.) \cite{OliveiraAlvesFerreira2013}.}
\label{fig: histogram}
\end{figure}

In 2D, like in 1D, the height field exhibits a linear growth in time with average velocity $v_\infty$, over which fluctuations develop with a $t^{\beta}$ scaling. All 2D KPZ numerical works  \cite{HalpinHealy2012, OliveiraAlvesFerreira2013} find that the distribution of the reduced fluctuations $\tilde h$ are non-Gaussian, and suggest, like in 1D, the existence of different geometry-dependent universality subclasses. However, the corresponding distribution functions have no known analytical form like Tracy-Widom or Baik-Rains distributions in 1D. To investigate the statistical properties of the phase fluctuations, we record $5120$ independent realisations of the time evolution of $\psi$, from which we extract the phase and unwrap it in time at a fixed space point. The presence of a few space-time vortex-anti-vortex pairs with random core positions induces phase slips at the position of interest, which are non-instantaneous phase jumps close to (but lower than) $2\pi$.

As already evidenced in 1D \cite{Fontaine2021}, these phase slips result in the existence of replicas of the distribution of the fluctuations separated by approximately $2\pi$. As detailed in Appendix \ref{app: phase}, one can select the main central distribution at each time instants in the KPZ window and sum them after appropriate re-scaling. The obtained distribution is shown in Fig.~\ref{fig: histogram}. It is markedly distinct from a Gaussian distribution of same mean and variance, and matches the distribution \footnote{In Ref. \cite{OliveiraAlvesFerreira2013}, the numerical data for the distribution of the fluctuations of the 2D KPZ interface is fitted using a Gumbel distribution with given parameters. Thus, in Fig.~\ref{fig: histogram} we used the same Gumbel function, which is equivalent to comparing with the KPZ numerical data on the range probed in our work. However, let us note that  the Gumbel distribution cannot be expected to be the true distribution for the 2D KPZ, and it should not be used to model the far tails. Since these tails are not resolved in our data, the comparison with the Gumbel function is appropriate.} computed in 2D KPZ large-scale numerical simulations for flat interfaces \cite{HalpinHealy2012, OliveiraAlvesFerreira2013}, which is the expected universality sub-class for polaritons in the absence of confinement potential \cite{Deligiannis2021}. This result for the fluctuations distribution enforces the evidence for KPZ universality in 2D polariton condensates.

\section{Conclusions and perspectives}

In this work, we have shown, using numerical simulations of the discrete stochastic generalised Gross-Pitaevskii equation, that a KPZ regime can be achieved in discrete 2D driven-dissipative exciton polariton condensates  under incoherent pumping. We have obtained the condensate spatiotemporal first-order correlation function, and shown that, for the parameters studied, it exhibits stretched exponential behavior, both in space and time, with critical exponents characteristic of the 2D KPZ universality class. This scaling persists in a finite region of pump strength and grid spacing. Moreover, the associated scaling function accurately matches with the theoretical KPZ scaling function given by functional Renormalization Group methods. We have also obtained the distribution of the phase fluctuations, which is highly non-Gaussian and very close to the distribution computed in numerical simulations of KPZ interfaces. This is a compelling evidence that the phase fluctuations behave as the KPZ stochastic process.

Our findings open promising perspectives. On the theoretical side, it would be desirable to obtain a complete description of the phase diagram of a polariton condensate in 2D, with a clear understanding of the interplay between the various possible regimes, as {\it e.g.} non-equilibrium BKT, and KPZ. This remains a challenge as the parameter space to explore is very large. On the experimental side, the demonstration of the KPZ scaling in a 1D polariton condensate was recently realized in a lattice, where real space was discretized~\cite{Fontaine2021}. Since our simulations predict that this is the optimal situation to prevent topological defect proliferation in 2D, similar techniques could be used in 2D. The evidence of a KPZ regime in a 2D polariton condensate would constitute a major breakthrough, especially since a convincing experimental realization of KPZ universality class in 2D is still missing and actively sought for. 

\begin{acknowledgments}

We acknowledge stimulating discussions with Iacopo Carusotto. K.D. acknowledges the European Union Horizon 2020 research and innovation programme under the Marie Sk\l{}odowska-Curie grant agreement No 754303. L.C. acknowledges support from the French ANR through the project NeqFluids (grant ANR-18-CE92-0019) and support from Institut Universitaire de France (IUF).  Q.F., S.R., and J.B. acknowledge support from the Paris Ile-de-France Région in the framework of DIM SIRTEQ, the French RENATECH network, the H2020-FETFLAG project PhoQus (820392), the QUANTERA project Interpol (ANR-QUAN-0003-05), the European Research Council via the project ARQADIA (949730).

\end{acknowledgments}

\appendix

\section{Discrete gGPE,  numerical integration, detection of vortices}
\label{app: vortices}

Our work is based on the 
numerical solution of  the generalized Gross-Pitaevskii equation~(\ref{eq: model}) by discretizing it on a grid of spacing $dx$. This may describe both a continuous system, by taking the limit of lattice spacing tending to zero, as well as a specific lattice geometry, which could be realized in the polariton experiment. We provide in this Appendix some further information
concerning the mapping to the lattice model, the numerical integration and the method for identifying the vortices.

\subsection{Derivation of the lattice model}

In order to describe the case of polaritons on a lattice, we start from Eq.~(\ref{eq: model}) 
\begin{align}
i \hbar\partial_t \psi = &\Bigg[ - \left(\frac{\hbar^2}{2m} -i \dfrac{ \hbar\gamma_{l,2}}{2} \right)\nabla^2 + V_\ell \nonumber \\ 
& +2g_r \dfrac{P}{\gamma_r \left( 1 + {\left\vert \psi \right \vert^2}/{n_s} \right)} + g \left\lvert \psi \right\rvert^2 \nonumber  \\
& +i\frac{\hbar \gamma_{l,0}}{2} \left( \dfrac{p}{1 + {\left\vert \psi \right\vert^2}/{n_s}} -1 \right) \Bigg] \psi + \hbar \xi \,,
\label{eq:model-lattice}
\end{align} 

where $V_\ell(\vx)$ is the periodic lattice potential. In the deep lattice regime,  the condensate wavefunction can be expanded on the Wannier function basis $\psi(t,{\vx})=\sum_{\vj} w_{\vj}({\vx}) \psi_{\vj}(t)$, with $w_{\vj}({\vx})=w(\vx-\vx_j)$ where $w(\vx)$ is localized around $\vx=0$. The Wannier functions are a set of real, localized, orthonormal wavefunctions centered on the lattice sites $\vx_{\vj}$, where ${\vj}=\{j_x,j_y\}$ identifies each lattice site of the 2D grid. 

We shall assume in the following for simplicity that the lattice potential is separable in the $x,y$ directions, \textit{i.e.} can be written as 
$V_\ell(\vx)=V(x)+V(y)$, and we denote by $dx$ the lattice spacing. Simple examples of $V(x)$ are the sinusoidal case $V(x)=V_0 \sin^2(k_0 x)$, with $k_0= \pi/dx$ or the Kronig-Penney potential $V(x)=V_0 \sum_n \Theta(x-(n dx +b))\Theta((n+1) dx -x)$, with $b$ being a parameter associated to the width of the potential barrier and $\Theta(x)$ the Heaviside step function. If the potential is separable, one also has $w({\vx})=w_{1d}(x) w_{1d}(y)$ with $w_{1d}(x)$ the Wannier function for the one-dimensional lattice problem. We will further assume that $|\psi|^2/n_s\ll 1$ such that the terms involving the pump can be Taylor expanded to first order. This hypothesis may be easily released by treating in an analogous fashion the higher-order  terms of the Taylor expansion. 

Under the above assumptions, by projecting Eq.~(\ref{eq:model-lattice}) onto the Wannier function basis and by neglecting overlaps among Wannier functions on different sites in the non-linear terms, 
we obtain  the following discrete non-linear Schr\"odinger equation
\begin{align}
    i \hbar  \partial_t \psi_{\vj} =& - K ( \psi_{j_x+1,j_y} + \psi_{j_x-1,j_y} +\psi_{j_x,j_y+1} + \psi_{j_x,j_y-1} ) \nonumber \\ 
    &+ W_0  \psi_{\vj} + \left( U- 2PU_r/(\gamma_r n_s) \right) |\psi_{\vj}|^2 \psi_{\vj}  \nonumber \\ 
    &+ i \frac{{\gamma}_{l,0}}{2} (p-1 - p |\psi_{\vj}|^2/{\cal N}_s ) \psi_{\vj} + \hbar \xi_{\vj}
    \label{discr-GPE-app}
\end{align}
where the parameters entering in Eq.~(\ref{discr-GPE-app}) are
\begin{align}
K&= \int dx \,w_{1d}(x) \Bigg[ \Big(\frac{\hbar^2}{2m} - i \frac { \hbar \gamma_{l,2}}{2}\Big)\partial^2_x \nonumber\\
 & - V(x)\Bigg] w_{1d}(x-dx)\nonumber\\
W_0 & =\int d^2\vx \,w({\vx}) \Bigg[ -\Big(\frac{\hbar^2 }{2m} - i \frac   {\hbar \gamma_{l,2}}{2} \Big) \nabla^2 + V_\ell({\vx})\Bigg] w({\vx})\nonumber\\
&+ 2 g_r P/\gamma_r\, ,
\end{align}
with $U=g \int d^2\vx [w({\vx})]^4 $, $U_r=g_r \int d^2\vx [w({\vx})]^4$, and ${\cal N}_s=n_s/\int d^2\vx [w({\vx})]^4$. The discretized noise is $ \xi_{\vj}(t)=\int d^2\vx w_{\vj}({\vx}) \xi(t,{\vx})$. In the numerical simulations, in order to make link with the continuous limit expressions, we have resummed the term $(p(1- |\psi_{\vj}|^2/{\cal N}_s )$  onto $p/(1+|\psi_{\vj}|^2/{\cal N}_s )$. Also, for convenience, we have chosen the zero of the energy such that  $W_0=4 K$.

The continuous limit of Eq.~(\ref{discr-GPE-app}) can be formally defined by the limit $\psi(\vx_j) = \lim_{dx\rightarrow0} \psi_{\vj}/dx$.
This allows one to relate the parameters of the lattice model to the ones of the effective continuous model, for example Re$(K)=\hbar^2/(2 m^* dx^2)$, with $m^*$ an effective mass. However, this limiting procedure assumes that the parameters entering the discrete GPE are independent from the lattice spacing, which is --strictly speaking-- not the case since the lattice potential enters in the definition of $K$. This implies that the effective mass will be slightly modified when changing $dx$.

\subsection{Details of the numerical integration}
The integration of the discretized Eq.~(\ref{eq: model}) is performed using the split-step Fourier method. This scheme consists of treating the kinetic and potential/interaction parts of the equation separately when propagating in time. More specifically, after suitable discretization in space, the wavefunction after a single time step $dt$ is constructed by the following three building blocks: i) propagation by $dt$ in real space using the potential and interaction parts, ii) propagation of $dt$ in momentum space by the kinetic part, and iii) addition of  the stochastic contribution (Wigner noise). This procedure is repeated in a loop until the desired final time is reached. We note that, switching back and forth from real to momentum space is done particularly efficiently via the Fast Fourier Transform procedure. The source code of our solver is available at \href[]{https://github.com/kdeligia/2D_KPZ_polaritons}{this repository}.

\subsection{Methods for identifying the vortices}
In order to search for vortices in the phase of the condensate and locate their core, one usually computes the circulation $I = \oint_C \vnabla \theta \cdot d\vl$ on closed contours $C$ enclosing each point of the 2D grid. If a quantized vortex is present at a point, then $I = \kappa 2\pi$, where $\kappa$ is the vortex charge, else $I=0$ \cite{LarsonEdwards}. The closed contour $C$ should be defined such that it encloses a small part of the fluid containing up to one vortex core, in which case the circulation does not depend on the precise form of the contour.

However, this method quickly becomes intractable when applied to each point in a grid consisting of many points, such as a space grid with fine spatial discretization, or a space-time grid, defined by selecting one spatial direction and time. In the latter case, the time discretization is typically chosen as much finer than the spatial one, thus requiring an extremely large number of points in order to study the dynamics of the condensate until some time after the steady state has been reached. We circumvent this issue by computing the curl of the gradient of the phase of the condensate, $\vnabla \times \vnabla \theta$. By definition, the projection of this quantity in each of the three directions, the two spatial and the temporal ones, is related to the infinitesimal circulation of the gradient of the phase around each point of the grid,
\begin{equation}
\vnabla \times \vnabla \theta \cdot \vec{S} \rightarrow I \,,
\end{equation}
where $\vec{S}$ is the normal vector,
\begin{equation}
\vec{S} = dx^2 \hat{t} + dx dt \hat{x} + dx dt \hat{y} \,,
\end{equation}
 $dx$ is the unit length, which is chosen equal for both the $x$ and $y$ directions, and $dt$ is the unit time.

From our numerical simulations, we extract the phase of the condensate in the $(x,y)$ plane in an appropriate time window in the steady state, thus effectively creating a 3D dataset. We then compute the components of the gradient numerically along each direction separately, once the phase has been unwrapped in that direction. Finally, we compute each of the components of the curl straightforwardly. Overall, this procedure allows us to identify, in a particularly efficient manner, both the spatial vortices, as well as the space-time topological defects, by mapping the $2+1$-dimensional problem to three distinct $2$-dimensional ones.

In order to compute the vortex density, we study each component of the curl separately. More specifically, for each component $\lbrace{x, y, t \rbrace}$, we count the number of non-zero values of each curl component for each fixed ``slice" $\lbrace{x_i, y_i, t_i \rbrace}$, which allows us to compute the mean number of defects,
\begin{subequations}
\begin{align}
\overline{N}_{\text{vort}, xy} &= \dfrac{1}{\#{t_i}} \sum_{t_i \in \Delta t} \text{count}_{\text{vort}}(t_i)\,, \\
\overline{N}_{\text{vort}, xt} &= \dfrac{1}{\#{y_i}} \sum_{y_i} \text{count}_{\text{vort}}(y_i)\,, \\
\overline{N}_{\text{vort}, yt} &= \dfrac{1}{\#{x_i}} \sum_{x_i} \text{count}_{\text{vort}}(x_i)\,,
\end{align}
\end{subequations}
where $\Delta t=100ps$ starting from an initial sampling time $t_0=100ps$. We then normalize our result by dividing with the total number of points in the $2D$ grid. We present our results for the mean occupation of the grid by topological defects in Figs.~\ref{fig: fig1} and \ref{fig: fig2}.
In all our simulations, we observed that the size of the vortex cores is smaller or of the order of the spatial discretization $dx$. This is compatible with the order of magnitude of the typical length-scale of our model, which can be estimated by comparing the dissipative part of the kinetic term with the gain term (since the interactions are zero). One finds that it is proportional to the ratio $\sqrt{\gamma_{l,2}/\gamma_{l,0}}$ and to the relative density fluctuations, and is hence expected to be smaller than $dx$.

\section{Robustness of the KPZ regime with respect to the pump power}
\label{app: robust}

\begin{figure}[ht!]
\includegraphics[scale=1]{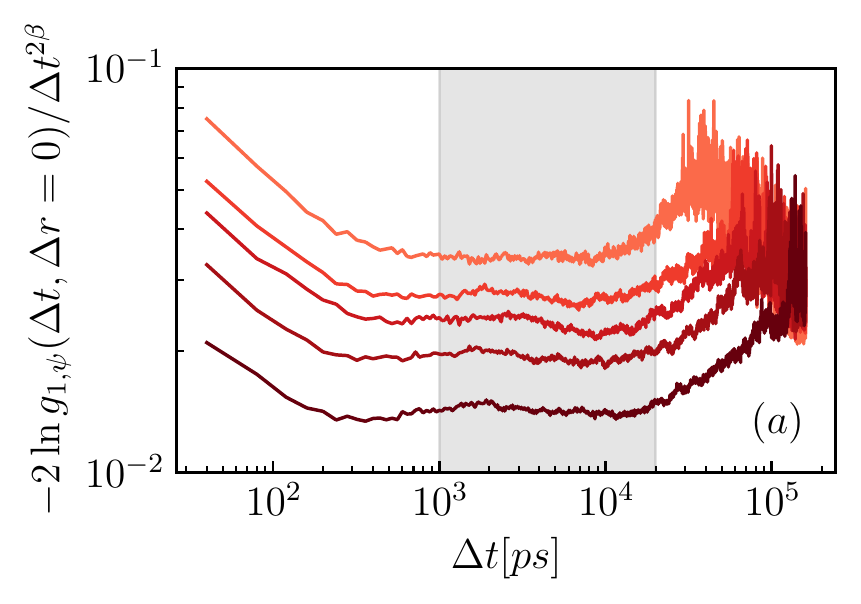}
\includegraphics[scale=1]{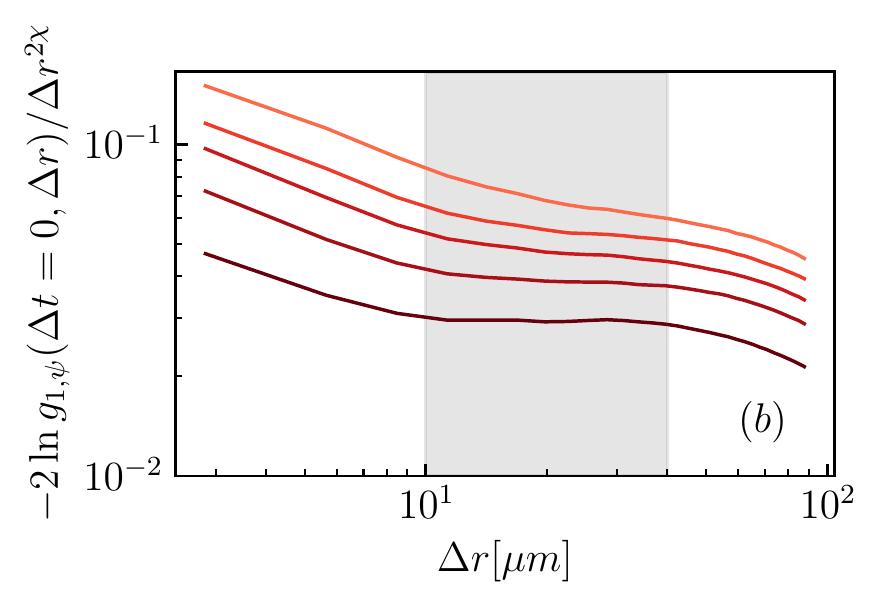}
\caption{(a) Equal-space and (b) equal-time correlation functions $g_{1, \psi}$, with darker colours corresponding to larger values of the pump parameter, $p=1.6, 1.7, 1.8, 2, 2.5$. The correlations are compensated by the appropriate KPZ power laws, such that the corresponding data appear as plateaus in the KPZ spatial and temporal ranges, which are highlighted in gray.}
\label{fig: correlations_scan}
\end{figure}

In order to test the robustness of the KPZ regime, we compute the  correlation function $g_{1,\psi}$ for different values of the reduced pump $p$. The equal-time and equal-space correlations, compensated by the KPZ power-laws, are displayed in Fig.~\ref{fig: correlations_scan}. We find that the temporal scaling is very robust, as plateaus for time delays spanning approximately $10^3 - 10^4 ps$ are clearly identified for all values of the pump. The spatial scaling turns out to be more sensitive to the pump  variations. Plateaus for spatial separations approximately $10 - 40 \mu m$ are apparent for $p\gtrsim 1.8$, but not for lower values of the pump. The progressive departure from the KPZ regime visible in the spatial scaling can be attributed to the increasing effect of density-phase correlations, as shown below. To summarize, we find a robust KPZ scaling for the correlation functions in the moderately-high pump regime, especially stable for the temporal behavior.

As expounded in the main text, the connection between the correlations of the condensate field $\psi$ and the correlations of the phase field $\theta$ itself relies on some assumptions. In order to check their validity, we extract the density $n(t, \vrr)$ and phase $\theta(t, \vrr)$ from the condensate field, and we compute separately the contributions to $g_{1,\psi}$ from the density correlations
\begin{equation}
g_{1, n}(\Delta t, \Delta r) = \frac{\left\langle \sqrt{n(t_0, \vrr_0) n(t_0 + \Delta t, \vrr_0 + \Delta \vrr)} \right\rangle}{\sqrt{ \left\langle n(t_0, \vrr_0) \right\rangle }\sqrt{\left\langle n(t_0 + \Delta t, \vrr_0 + \Delta \vrr) \right\rangle}}
\end{equation}
and the one from the phase correlations
\begin{equation}
g_{1, \theta}(\Delta t, \Delta r) = \left\langle e^{i \left(\theta(t_0 + \Delta t, \vrr_0 + \Delta \vrr) -\theta(t_0, \vrr_0)\right) } \right\rangle\,.
\end{equation}

\begin{figure}[ht!]
\includegraphics[scale=1]{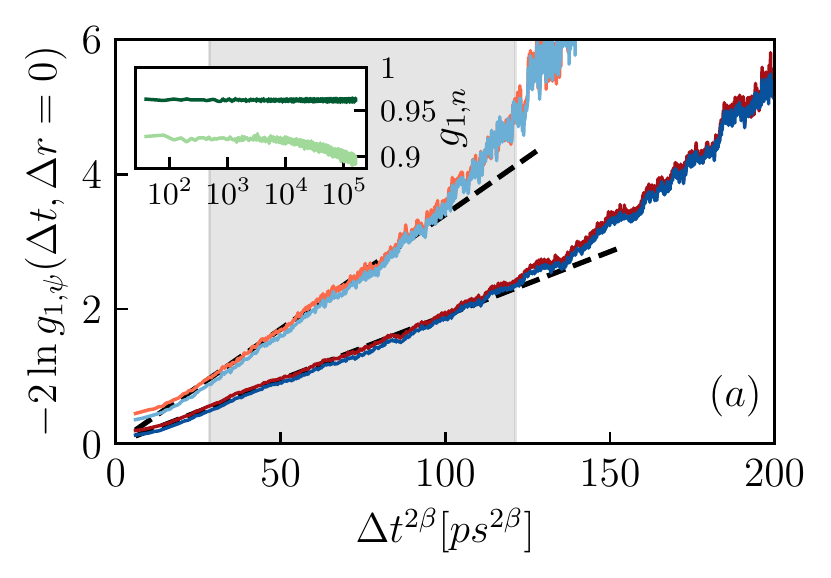}
\includegraphics[scale=1]{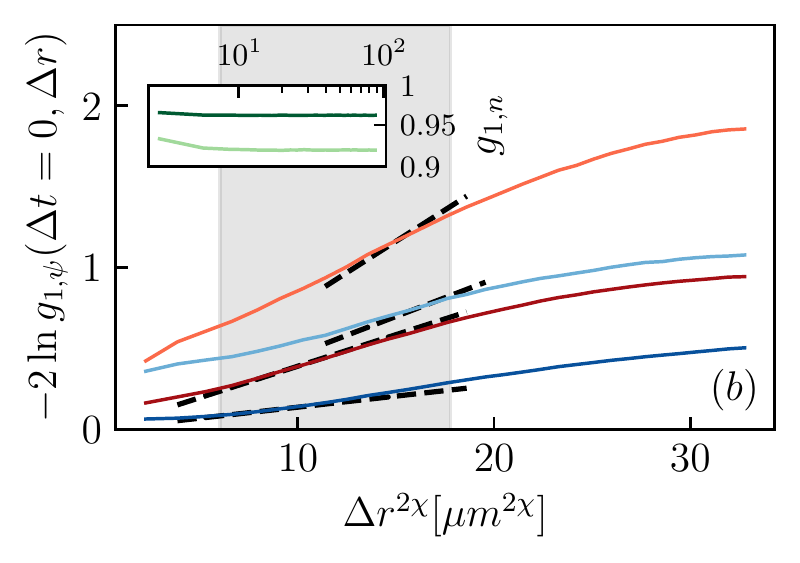}
\caption{(a) Equal-space and (b) equal-time correlation functions $g_{1, \psi}$ (red) and $g_{1, \theta}$ (blue), for $p=1.6$ (lighter shade) and $p=2$ (darker shade). The dashed lines indicate the corresponding stretched exponentials with KPZ exponents. In panel (b), the lighter shades are shifted upwards after multiplying by a factor of $1.25$ in order to avoid overlapping with the darker shades. Inset: Density-density correlation $g_{1, n}$ for these values of the pump.}
\label{fig: correlations_scan_extra}
\end{figure}

The results for $p=2$ and $p=1.6$ are displayed in Fig.~\ref{fig: correlations_scan_extra}. We observe that $g_{1,n}$ is very close to unity and constant over the whole KPZ spatial and temporal ranges for both pump powers. Thus, the role of density-density correlations is indeed negligible. The density-phase time correlations are almost vanishing, since $g_{1,\psi}(\Delta t, 0)$ and $g_{1,\theta}(\Delta t, 0)$ almost perfectly coincide over the whole KPZ time range, as shown in panel (a) of Fig.~\ref{fig: correlations_scan_extra}. The density-phase space correlations are more important, and their effect increases while decreasing the pump. They are shown in panel (b) of Fig.~\ref{fig: correlations_scan_extra}. For  $p=2$, the KPZ stretched exponential well models the behavior of both $g_{1,\psi}$ and $g_{1,\theta}$. We checked that the difference between the two curves is just a multiplicative factor. Hence, the density-phase correlations are almost constant in the spatial KPZ window, and the scaling of $g_{1,\psi}$ follows the scaling of the phase-phase correlations, as expected in a KPZ regime.
For $p=1.6$, the KPZ stretched exponential no longer provides a good fit for the behavior of $g_{1,\psi}$ (as already visible in Fig.~\ref{fig: correlations_scan_extra}) and neither for $g_{1,\theta}$. Moreover the density-phase correlations start varying within the KPZ window, the difference between  $g_{1,\psi}$ and $g_{1,\theta}$ is no longer a mere multiplicative factor. Thus, one concludes that for $p=2$, the behavior of the correlations $g_{1,\psi}$ indeed reflects the KPZ scaling of the phase itself, both in space and time, whereas for $p=1.6$, the  density-phase space correlations also contribute to the behavior of the  $g_{1,\psi}$, and the KPZ regime is hindered in space, although it persists in time.

\section{KPZ scaling function}
\label{app: Frond}

In the functional Renormalization Group analysis of Ref.~\cite{KlossCanetWschebor2012}, the following scaling function $\mathring{F}$ is computed
\begin{equation}
\bar{C}(\omega, \vp) =  \frac{2}{|\vp|^{d+2+\chi}} \bar{C}_0 \mathring{F}\left(\bar{y}_0 \frac{\omega}{|\vp|^z}\right)\,,\label{eq: Cscal}
\end{equation}
where $\bar{C}_0$, $\bar{y}_0$ are non-universal normalization constants, and $\bar{C}(\omega, \vp)$ is the Fourier transform of the connected correlation function defined in Eq.~\eqref{eq: defCh} of the main text, that is
\begin{equation}
C(t,\vrr) =  \int_{-\infty}^{\infty} \! \frac{d \omega}{2\pi} \! \int \! \! \frac{d^d \vp} {(2\pi)^d}\left(e^{-i(\omega t-\vp\cdot\vrr)}-1\right)\bar{C}(\omega,\vp)\, .
\label{eq: c1fourier}
\end{equation}
To compute the scaling function $F$ defined in Eq.~\eqref{eq: scaling} of the main text,
one can replace $\bar{C}(\omega,\vp)$ by its scaling form \eqref{eq: Cscal}, switch to polar coordinates, and change frequency variable to $\tau = \bar{y}_0 \omega/ p^z$. One obtains in 2D
\begin{align}
C(t,\vrr) &= \frac{\bar{C_0}}{\bar{y}_0}\frac{1}{2\pi^2}\int_{0}^{\infty} \! d\tau \mathring{F}(\tau)\int_{0}^{\infty}  \frac{dp}{p^{3+\chi-z}} \nonumber\\
&\times \Big[\cos(p^z \tau t/\bar{y}_0) {\cal B}_J(0, p r) -1\Big]
\end{align}
where ${\cal B}_J$ is a Bessel function, and the parity of $\mathring{F}$ has been used. By changing momentum variable to $u=p^z \tau t/\bar{y}_0$, one finally obtains
\begin{align}
C(t,\vrr) &= C_0 t^{2\chi/z} F(y_0 x/t^{1/z}) \nonumber\\
F(y) &= \int_{0}^{\infty} \! d\tau \mathring{F}(\tau)\tau^{2\chi/z}\int_{0}^{\infty}\frac{du}{u^{2\chi/z}}\label{eq: F}\\
& \times\Big[\cos(u) {\cal B}_J(0, (u/\tau)^{1/z} y) -1\Big] \nonumber\\
C_0 &= \frac{\bar{C_0}}{(2\pi^2)z\bar{y}_0^{1+2\chi}} \nonumber\\
y_0 &= \bar{y_0}^{1/z}
\end{align}
The normalization constants $C_0$ and $y_0$ are not universal, and have to be prescribed. In this work, we fix $C_0$ and $y_0$ such that $F_0 = 1$ and $F_\infty = 0.45$. The integrals in Eq.~(\ref{eq: F}) were computed numerically, using the tabulated data for $\mathring{F}$ from Ref.~\cite{KlossCanetWschebor2012}.

\section{Phase distribution}
\label{app: phase}

\begin{figure}[ht!]
\includegraphics[scale=1]{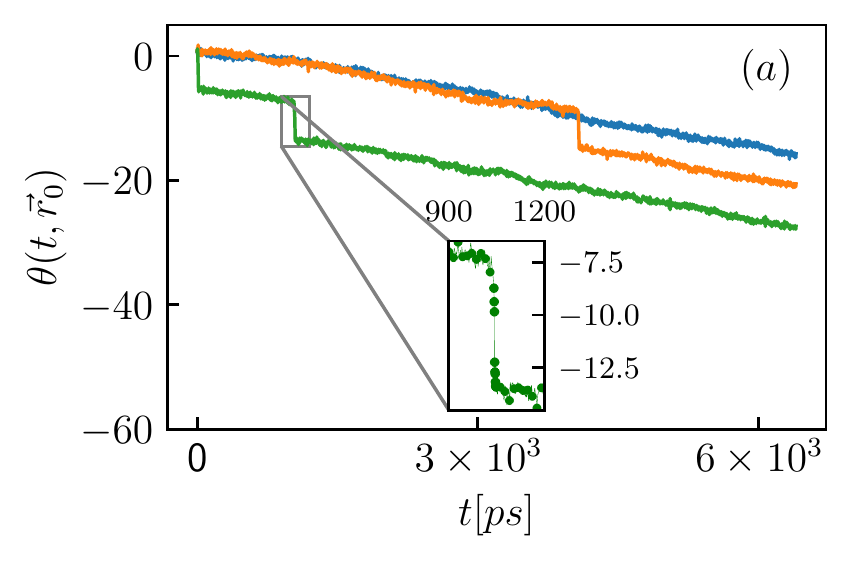}
\includegraphics[scale=1]{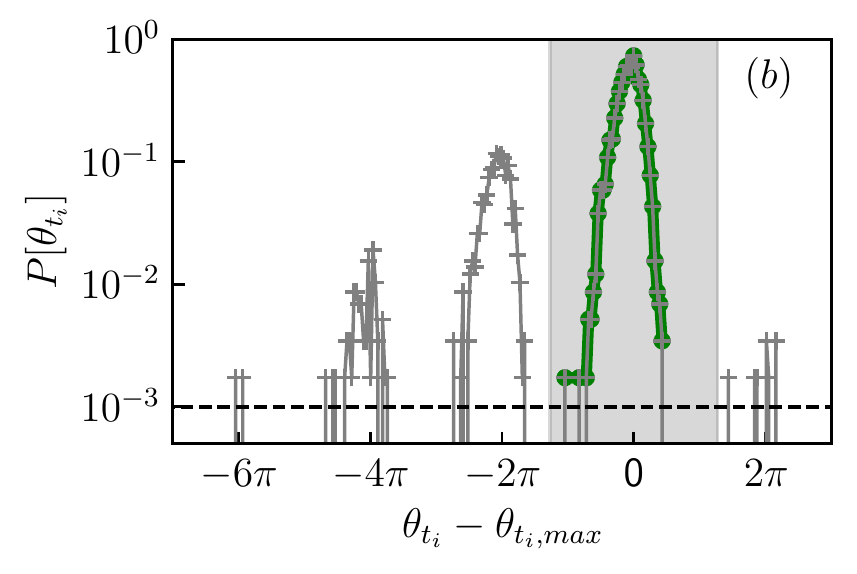}
\caption{(a) Three typical unwrapped phase trajectories, displaying zero, one, or two  phase jumps. (b) $P[\theta(t_i)]$ for a given $t_i$ (gray symbols). We select the data for which $\left\lvert \theta_{t_i} - \theta_{t_i, \textrm{max}} \right\rvert < \epsilon$, and then we perform a secondary selection for $P[\theta_{t_i}] > P_{\textrm{min}}$, where $P_{\textrm{min}}$ is shown as a dashed black line. The data which satisfy both criteria are highlighted (green dots).}
\label{fig: treating}
\end{figure}

As mentioned in the main text, the occurrence of a few topological defects, which are space-time vortex-anti-vortex pairs, generates phase slips of the time-unwrapped phase when crossing a pair. As can be seen in the typical unwrapped phase trajectories shown in panel (a) of Fig.~\ref{fig: treating}, these phase slips consist of abrupt but non-instantaneous changes of the phase of approximately $2\pi$. The existence of these jumps disrupts the leading linear behavior in time, and thus prevents from consistently defining the reduced fluctuations as $\delta \theta \propto (\theta - \left\langle \theta \right\rangle)/t^{\beta}$.

In order to circumvent this issue, we compute the probability density function $P[\theta_{t_i}]$ for each discrete time instant $t_i$ in the KPZ window. An example of such a distribution is shown in panel (b) of Fig.~\ref{fig: treating}. Similarly to what was reported in 1D \cite{Fontaine2021}, the phase slips give rise to secondary peaks in the distribution of the phase. The fluctuations of the realization where no, one, two, \dots \;jumps occurred populate the distribution in the first, second, third, \dots \;peak, where the different peaks are separated by approximately $2\pi$. These peaks have a similar shape, which shows that the dynamics is in fact piece-wise KPZ in between the jumps. A similar behavior was also reported in \cite{Fontaine2021} in 1D and shown not to disrupt the emergence of KPZ universality. To study the shape of the main distribution with more accuracy, we concentrate on the fluctuations in the first peak for each time instant, by requiring $\left\lvert \theta_{t_i} - \theta_{t_i, \textrm{max}} \right\rvert < \epsilon$, where $\theta_{t_i, \textrm{max}}$ is the point for which $P[\theta_{t_i, \textrm{max}}] = \textrm{max}(P[\theta_{t_i}])$, and we focus on the subset of them where $P[\theta_{t_i}] > P_{\textrm{min}}$. For each time instant, the corresponding distribution is centered at zero, by subtracting the first moment of the selected data, $\delta \theta_{t_i} = \theta_{t_i} - \langle\theta_{t_i}\rangle$, and normalized to unit variance. We then sum the fluctuations of the first peaks, properly normalized, for all time instants in the appropriate KPZ time window. 

Let us mention that we have also performed a detailed analysis of the skewness and excess kurtosis of the phase fluctuations. These quantities are insensitive to normalisation issues, whereas the data analysis of the distribution can suffer from unavoidable errors due to the empirical choice of $\epsilon$ and $P_{\textrm{min}}$, and the small number of discrete bins for given $t_i$. This analysis is reported in \cite{konstantinos_thesis}, and confirms that the skewness and excess kurtosis computed from our data is in good agreement with the universal values obtained from KPZ numerical simulations \cite{HalpinHealy2012, OliveiraAlvesFerreira2013}.


\begin{thebibliography}{74}%
\makeatletter
\providecommand \@ifxundefined [1]{%
 \@ifx{#1\undefined}
}%
\providecommand \@ifnum [1]{%
 \ifnum #1\expandafter \@firstoftwo
 \else \expandafter \@secondoftwo
 \fi
}%
\providecommand \@ifx [1]{%
 \ifx #1\expandafter \@firstoftwo
 \else \expandafter \@secondoftwo
 \fi
}%
\providecommand \natexlab [1]{#1}%
\providecommand \enquote  [1]{``#1''}%
\providecommand \bibnamefont  [1]{#1}%
\providecommand \bibfnamefont [1]{#1}%
\providecommand \citenamefont [1]{#1}%
\providecommand \href@noop [0]{\@secondoftwo}%
\providecommand \href [0]{\begingroup \@sanitize@url \@href}%
\providecommand \@href[1]{\@@startlink{#1}\@@href}%
\providecommand \@@href[1]{\endgroup#1\@@endlink}%
\providecommand \@sanitize@url [0]{\catcode `\\12\catcode `\$12\catcode
  `\&12\catcode `\#12\catcode `\^12\catcode `\_12\catcode `\%12\relax}%
\providecommand \@@startlink[1]{}%
\providecommand \@@endlink[0]{}%
\providecommand \url  [0]{\begingroup\@sanitize@url \@url }%
\providecommand \@url [1]{\endgroup\@href {#1}{\urlprefix }}%
\providecommand \urlprefix  [0]{URL }%
\providecommand \Eprint [0]{\href }%
\providecommand \doibase [0]{https://doi.org/}%
\providecommand \selectlanguage [0]{\@gobble}%
\providecommand \bibinfo  [0]{\@secondoftwo}%
\providecommand \bibfield  [0]{\@secondoftwo}%
\providecommand \translation [1]{[#1]}%
\providecommand \BibitemOpen [0]{}%
\providecommand \bibitemStop [0]{}%
\providecommand \bibitemNoStop [0]{.\EOS\space}%
\providecommand \EOS [0]{\spacefactor3000\relax}%
\providecommand \BibitemShut  [1]{\csname bibitem#1\endcsname}%
\let\auto@bib@innerbib\@empty
\bibitem [{\citenamefont {Bak}\ \emph {et~al.}(1987)\citenamefont {Bak},
  \citenamefont {Tang},\ and\ \citenamefont
  {Wiesenfeld}}]{BakTangWiesenfeld1987}%
  \BibitemOpen
  \bibfield  {author} {\bibinfo {author} {\bibfnamefont {P.}~\bibnamefont
  {Bak}}, \bibinfo {author} {\bibfnamefont {C.}~\bibnamefont {Tang}},\ and\
  \bibinfo {author} {\bibfnamefont {K.}~\bibnamefont {Wiesenfeld}},\ }\bibfield
   {title} {\bibinfo {title} {Self-organized criticality: {A}n explanation of
  the ${1/f}$ noise},\ }\href {https://doi.org/10.1103/PhysRevLett.59.381}
  {\bibfield  {journal} {\bibinfo  {journal} {Phys. Rev. Lett.}\ }\textbf
  {\bibinfo {volume} {59}},\ \bibinfo {pages} {381} (\bibinfo {year}
  {1987})}\BibitemShut {NoStop}%
\bibitem [{\citenamefont {Halpin-Healy}\ and\ \citenamefont
  {Zhang}(1995)}]{HalpinHealyZhang1995}%
  \BibitemOpen
  \bibfield  {author} {\bibinfo {author} {\bibfnamefont {T.}~\bibnamefont
  {Halpin-Healy}}\ and\ \bibinfo {author} {\bibfnamefont {Y.-C.}\ \bibnamefont
  {Zhang}},\ }\bibfield  {title} {\bibinfo {title} {Kinetic roughening
  phenomena, stochastic growth, directed polymers and all that. aspects of
  multidisciplinary statistical mechanics},\ }\href
  {https://doi.org/https://doi.org/10.1016/0370-1573(94)00087-J} {\bibfield
  {journal} {\bibinfo  {journal} {Physics Reports}\ }\textbf {\bibinfo {volume}
  {254}},\ \bibinfo {pages} {215} (\bibinfo {year} {1995})}\BibitemShut
  {NoStop}%
\bibitem [{\citenamefont {Krug}(1997)}]{Krug1997}%
  \BibitemOpen
  \bibfield  {author} {\bibinfo {author} {\bibfnamefont {J.}~\bibnamefont
  {Krug}},\ }\bibfield  {title} {\bibinfo {title} {Origins of scale invariance
  in growth processes},\ }\href {https://doi.org/10.1080/00018739700101498}
  {\bibfield  {journal} {\bibinfo  {journal} {Advances in Physics}\ }\textbf
  {\bibinfo {volume} {46}},\ \bibinfo {pages} {139} (\bibinfo {year}
  {1997})}\BibitemShut {NoStop}%
\bibitem [{\citenamefont {Kardar}\ \emph {et~al.}(1986)\citenamefont {Kardar},
  \citenamefont {Parisi},\ and\ \citenamefont {Zhang}}]{KardarParisiZhang}%
  \BibitemOpen
  \bibfield  {author} {\bibinfo {author} {\bibfnamefont {M.}~\bibnamefont
  {Kardar}}, \bibinfo {author} {\bibfnamefont {G.}~\bibnamefont {Parisi}},\
  and\ \bibinfo {author} {\bibfnamefont {Y.}~\bibnamefont {Zhang}},\ }\bibfield
   {title} {\bibinfo {title} {Dynamic scaling of growing interfaces},\ }\href
  {https://doi.org/10.1103/PhysRevLett.56.889} {\bibfield  {journal} {\bibinfo
  {journal} {Phys. Rev. Lett.}\ }\textbf {\bibinfo {volume} {56}},\ \bibinfo
  {pages} {889} (\bibinfo {year} {1986})}\BibitemShut {NoStop}%
\bibitem [{\citenamefont {Takeuchi}(2018)}]{Takeuchi2018}%
  \BibitemOpen
  \bibfield  {author} {\bibinfo {author} {\bibfnamefont {K.~A.}\ \bibnamefont
  {Takeuchi}},\ }\bibfield  {title} {\bibinfo {title} {An appetizer to modern
  developments on the {K}ardar--{P}arisi--{Z}hang universality class},\ }\href
  {https://doi.org/https://doi.org/10.1016/j.physa.2018.03.009} {\bibfield
  {journal} {\bibinfo  {journal} {Physica A: Statistical Mechanics and its
  Applications}\ }\textbf {\bibinfo {volume} {504}},\ \bibinfo {pages} {77}
  (\bibinfo {year} {2018})}\BibitemShut {NoStop}%
\bibitem [{\citenamefont {Takeuchi}\ and\ \citenamefont
  {Sano}(2012)}]{TakeuchiSano2012}%
  \BibitemOpen
  \bibfield  {author} {\bibinfo {author} {\bibfnamefont {K.~A.}\ \bibnamefont
  {Takeuchi}}\ and\ \bibinfo {author} {\bibfnamefont {M.}~\bibnamefont
  {Sano}},\ }\bibfield  {title} {\bibinfo {title} {Evidence for
  geometry-dependent universal fluctuations of the {K}ardar-{P}arisi-{Z}hang
  interfaces in liquid-crystal turbulence},\ }\href
  {https://doi.org/10.1007/s10955-012-0503-0} {\bibfield  {journal} {\bibinfo
  {journal} {Journal of Statistical Physics}\ }\textbf {\bibinfo {volume}
  {147}},\ \bibinfo {pages} {853} (\bibinfo {year} {2012})}\BibitemShut
  {NoStop}%
\bibitem [{\citenamefont {Najem}\ \emph {et~al.}(2020)\citenamefont {Najem},
  \citenamefont {Krayem}, \citenamefont {Ala-Nissila},\ and\ \citenamefont
  {Grant}}]{Najem2020}%
  \BibitemOpen
  \bibfield  {author} {\bibinfo {author} {\bibfnamefont {S.}~\bibnamefont
  {Najem}}, \bibinfo {author} {\bibfnamefont {A.}~\bibnamefont {Krayem}},
  \bibinfo {author} {\bibfnamefont {T.}~\bibnamefont {Ala-Nissila}},\ and\
  \bibinfo {author} {\bibfnamefont {M.}~\bibnamefont {Grant}},\ }\bibfield
  {title} {\bibinfo {title} {Kinetic roughening of the urban skyline},\ }\href
  {https://doi.org/10.1103/PhysRevE.101.050301} {\bibfield  {journal} {\bibinfo
   {journal} {Phys. Rev. E}\ }\textbf {\bibinfo {volume} {101}},\ \bibinfo
  {pages} {050301} (\bibinfo {year} {2020})}\BibitemShut {NoStop}%
\bibitem [{\citenamefont {Kasprzak}\ \emph {et~al.}(2006)\citenamefont
  {Kasprzak}, \citenamefont {Richard}, \citenamefont {Kundermann},
  \citenamefont {Baas}, \citenamefont {Jeambrun}, \citenamefont {Keeling},
  \citenamefont {Marchetti}, \citenamefont {Szyma{\'n}ska}, \citenamefont
  {Andr{\'e}}, \citenamefont {Staehli}, \citenamefont {Savona}, \citenamefont
  {Littlewood}, \citenamefont {Deveaud},\ and\ \citenamefont
  {Dang}}]{Kasprzak2006}%
  \BibitemOpen
  \bibfield  {author} {\bibinfo {author} {\bibfnamefont {J.}~\bibnamefont
  {Kasprzak}}, \bibinfo {author} {\bibfnamefont {M.}~\bibnamefont {Richard}},
  \bibinfo {author} {\bibfnamefont {S.}~\bibnamefont {Kundermann}}, \bibinfo
  {author} {\bibfnamefont {A.}~\bibnamefont {Baas}}, \bibinfo {author}
  {\bibfnamefont {P.}~\bibnamefont {Jeambrun}}, \bibinfo {author}
  {\bibfnamefont {J.~M.~J.}\ \bibnamefont {Keeling}}, \bibinfo {author}
  {\bibfnamefont {F.~M.}\ \bibnamefont {Marchetti}}, \bibinfo {author}
  {\bibfnamefont {M.~H.}\ \bibnamefont {Szyma{\'n}ska}}, \bibinfo {author}
  {\bibfnamefont {R.}~\bibnamefont {Andr{\'e}}}, \bibinfo {author}
  {\bibfnamefont {J.~L.}\ \bibnamefont {Staehli}}, \bibinfo {author}
  {\bibfnamefont {V.}~\bibnamefont {Savona}}, \bibinfo {author} {\bibfnamefont
  {P.~B.}\ \bibnamefont {Littlewood}}, \bibinfo {author} {\bibfnamefont
  {B.}~\bibnamefont {Deveaud}},\ and\ \bibinfo {author} {\bibfnamefont {L.~S.}\
  \bibnamefont {Dang}},\ }\bibfield  {title} {\bibinfo {title} {Bose--einstein
  condensation of exciton polaritons},\ }\href
  {https://doi.org/10.1038/nature05131} {\bibfield  {journal} {\bibinfo
  {journal} {Nature}\ }\textbf {\bibinfo {volume} {443}},\ \bibinfo {pages}
  {409} (\bibinfo {year} {2006})}\BibitemShut {NoStop}%
\bibitem [{\citenamefont {Carusotto}\ and\ \citenamefont
  {Ciuti}(2013)}]{CarusottoCiuti2013}%
  \BibitemOpen
  \bibfield  {author} {\bibinfo {author} {\bibfnamefont {I.}~\bibnamefont
  {Carusotto}}\ and\ \bibinfo {author} {\bibfnamefont {C.}~\bibnamefont
  {Ciuti}},\ }\bibfield  {title} {\bibinfo {title} {Quantum fluids of light},\
  }\href {https://doi.org/10.1103/RevModPhys.85.299} {\bibfield  {journal}
  {\bibinfo  {journal} {Rev. Mod. Phys.}\ }\textbf {\bibinfo {volume} {85}},\
  \bibinfo {pages} {299} (\bibinfo {year} {2013})}\BibitemShut {NoStop}%
\bibitem [{\citenamefont {Bloch}\ \emph {et~al.}(2022)\citenamefont {Bloch},
  \citenamefont {Carusotto},\ and\ \citenamefont
  {Wouters}}]{BlochCarusottoWouters2022}%
  \BibitemOpen
  \bibfield  {author} {\bibinfo {author} {\bibfnamefont {J.}~\bibnamefont
  {Bloch}}, \bibinfo {author} {\bibfnamefont {I.}~\bibnamefont {Carusotto}},\
  and\ \bibinfo {author} {\bibfnamefont {M.}~\bibnamefont {Wouters}},\
  }\bibfield  {title} {\bibinfo {title} {Non-equilibrium bose–einstein
  condensation in photonic systems},\ }\bibfield  {journal} {\bibinfo
  {journal} {Nature Reviews Physics}\ }\href
  {https://doi.org/10.1038/s42254-022-00464-0} {10.1038/s42254-022-00464-0}
  (\bibinfo {year} {2022})\BibitemShut {NoStop}%
\bibitem [{\citenamefont {Gladilin}\ \emph {et~al.}(2014)\citenamefont
  {Gladilin}, \citenamefont {Ji},\ and\ \citenamefont
  {Wouters}}]{GladilinJiWouters2014}%
  \BibitemOpen
  \bibfield  {author} {\bibinfo {author} {\bibfnamefont {V.~N.}\ \bibnamefont
  {Gladilin}}, \bibinfo {author} {\bibfnamefont {K.}~\bibnamefont {Ji}},\ and\
  \bibinfo {author} {\bibfnamefont {M.}~\bibnamefont {Wouters}},\ }\bibfield
  {title} {\bibinfo {title} {Spatial coherence of weakly interacting
  one-dimensional nonequilibrium bosonic quantum fluids},\ }\href
  {https://doi.org/10.1103/PhysRevA.90.023615} {\bibfield  {journal} {\bibinfo
  {journal} {Phys. Rev. A}\ }\textbf {\bibinfo {volume} {90}},\ \bibinfo
  {pages} {023615} (\bibinfo {year} {2014})}\BibitemShut {NoStop}%
\bibitem [{\citenamefont {Ji}\ \emph {et~al.}(2015)\citenamefont {Ji},
  \citenamefont {Gladilin},\ and\ \citenamefont
  {Wouters}}]{JiGladilinWouters2015}%
  \BibitemOpen
  \bibfield  {author} {\bibinfo {author} {\bibfnamefont {K.}~\bibnamefont
  {Ji}}, \bibinfo {author} {\bibfnamefont {V.~N.}\ \bibnamefont {Gladilin}},\
  and\ \bibinfo {author} {\bibfnamefont {M.}~\bibnamefont {Wouters}},\
  }\bibfield  {title} {\bibinfo {title} {Temporal coherence of one-dimensional
  nonequilibrium quantum fluids},\ }\href
  {https://doi.org/10.1103/PhysRevB.91.045301} {\bibfield  {journal} {\bibinfo
  {journal} {Phys. Rev. B}\ }\textbf {\bibinfo {volume} {91}},\ \bibinfo
  {pages} {045301} (\bibinfo {year} {2015})}\BibitemShut {NoStop}%
\bibitem [{\citenamefont {Altman}\ \emph {et~al.}(2015)\citenamefont {Altman},
  \citenamefont {Sieberer}, \citenamefont {Chen}, \citenamefont {Diehl},\ and\
  \citenamefont {Toner}}]{AltmanSiebererChenDiehlToner2015}%
  \BibitemOpen
  \bibfield  {author} {\bibinfo {author} {\bibfnamefont {E.}~\bibnamefont
  {Altman}}, \bibinfo {author} {\bibfnamefont {L.~M.}\ \bibnamefont
  {Sieberer}}, \bibinfo {author} {\bibfnamefont {L.}~\bibnamefont {Chen}},
  \bibinfo {author} {\bibfnamefont {S.}~\bibnamefont {Diehl}},\ and\ \bibinfo
  {author} {\bibfnamefont {J.}~\bibnamefont {Toner}},\ }\bibfield  {title}
  {\bibinfo {title} {Two-dimensional superfluidity of exciton polaritons
  requires strong anisotropy},\ }\href
  {https://doi.org/10.1103/PhysRevX.5.011017} {\bibfield  {journal} {\bibinfo
  {journal} {Phys. Rev. X}\ }\textbf {\bibinfo {volume} {5}},\ \bibinfo {pages}
  {011017} (\bibinfo {year} {2015})}\BibitemShut {NoStop}%
\bibitem [{\citenamefont {He}\ \emph {et~al.}(2015)\citenamefont {He},
  \citenamefont {Sieberer}, \citenamefont {Altman},\ and\ \citenamefont
  {Diehl}}]{HeSiebererAltmanDiehl2015}%
  \BibitemOpen
  \bibfield  {author} {\bibinfo {author} {\bibfnamefont {L.}~\bibnamefont
  {He}}, \bibinfo {author} {\bibfnamefont {L.~M.}\ \bibnamefont {Sieberer}},
  \bibinfo {author} {\bibfnamefont {E.}~\bibnamefont {Altman}},\ and\ \bibinfo
  {author} {\bibfnamefont {S.}~\bibnamefont {Diehl}},\ }\bibfield  {title}
  {\bibinfo {title} {Scaling properties of one-dimensional driven-dissipative
  condensates},\ }\href {https://doi.org/10.1103/PhysRevB.92.155307} {\bibfield
   {journal} {\bibinfo  {journal} {Phys. Rev. B}\ }\textbf {\bibinfo {volume}
  {92}},\ \bibinfo {pages} {155307} (\bibinfo {year} {2015})}\BibitemShut
  {NoStop}%
\bibitem [{\citenamefont {Squizzato}\ \emph {et~al.}(2018)\citenamefont
  {Squizzato}, \citenamefont {Canet},\ and\ \citenamefont
  {Minguzzi}}]{SquizzatoCanetMinguzzi2018}%
  \BibitemOpen
  \bibfield  {author} {\bibinfo {author} {\bibfnamefont {D.}~\bibnamefont
  {Squizzato}}, \bibinfo {author} {\bibfnamefont {L.}~\bibnamefont {Canet}},\
  and\ \bibinfo {author} {\bibfnamefont {A.}~\bibnamefont {Minguzzi}},\
  }\bibfield  {title} {\bibinfo {title} {{K}ardar-{P}arisi-{Z}hang universality
  in the phase distributions of one-dimensional exciton polaritons},\ }\href
  {https://doi.org/10.1103/PhysRevB.97.195453} {\bibfield  {journal} {\bibinfo
  {journal} {Phys. Rev. B}\ }\textbf {\bibinfo {volume} {97}},\ \bibinfo
  {pages} {195453} (\bibinfo {year} {2018})}\BibitemShut {NoStop}%
\bibitem [{\citenamefont {Deligiannis}\ \emph {et~al.}(2020)\citenamefont
  {Deligiannis}, \citenamefont {Squizzato}, \citenamefont {Minguzzi},\ and\
  \citenamefont {Canet}}]{Deligiannis2021}%
  \BibitemOpen
  \bibfield  {author} {\bibinfo {author} {\bibfnamefont {K.}~\bibnamefont
  {Deligiannis}}, \bibinfo {author} {\bibfnamefont {D.}~\bibnamefont
  {Squizzato}}, \bibinfo {author} {\bibfnamefont {A.}~\bibnamefont
  {Minguzzi}},\ and\ \bibinfo {author} {\bibfnamefont {L.}~\bibnamefont
  {Canet}},\ }\bibfield  {title} {\bibinfo {title} {Accessing
  {K}ardar-{P}arisi-{Z}hang universality sub-classes with exciton polaritons},\
  }\href {https://doi.org/10.1209/0295-5075/132/67004} {\bibfield  {journal}
  {\bibinfo  {journal} {EPL (Europhysics Letters)}\ }\textbf {\bibinfo {volume}
  {132}},\ \bibinfo {pages} {67004} (\bibinfo {year} {2020})}\BibitemShut
  {NoStop}%
\bibitem [{\citenamefont {He}\ \emph {et~al.}(2017)\citenamefont {He},
  \citenamefont {Sieberer},\ and\ \citenamefont {Diehl}}]{HeSiebererDiehl2017}%
  \BibitemOpen
  \bibfield  {author} {\bibinfo {author} {\bibfnamefont {L.}~\bibnamefont
  {He}}, \bibinfo {author} {\bibfnamefont {L.~M.}\ \bibnamefont {Sieberer}},\
  and\ \bibinfo {author} {\bibfnamefont {S.}~\bibnamefont {Diehl}},\ }\bibfield
   {title} {\bibinfo {title} {Space-time vortex driven crossover and vortex
  turbulence phase transition in one-dimensional driven open condensates},\
  }\href {https://doi.org/10.1103/PhysRevLett.118.085301} {\bibfield  {journal}
  {\bibinfo  {journal} {Phys. Rev. Lett.}\ }\textbf {\bibinfo {volume} {118}},\
  \bibinfo {pages} {085301} (\bibinfo {year} {2017})}\BibitemShut {NoStop}%
\bibitem [{\citenamefont {Fontaine}\ \emph {et~al.}(2022)\citenamefont
  {Fontaine}, \citenamefont {Squizzato}, \citenamefont {Baboux}, \citenamefont
  {Amelio}, \citenamefont {Lema{\^\i}tre}, \citenamefont {Morassi},
  \citenamefont {Sagnes}, \citenamefont {Le~Gratiet}, \citenamefont {Harouri},
  \citenamefont {Wouters}, \citenamefont {Carusotto}, \citenamefont {Amo},
  \citenamefont {Richard}, \citenamefont {Minguzzi}, \citenamefont {Canet},
  \citenamefont {Ravets},\ and\ \citenamefont {Bloch}}]{Fontaine2021}%
  \BibitemOpen
  \bibfield  {author} {\bibinfo {author} {\bibfnamefont {Q.}~\bibnamefont
  {Fontaine}}, \bibinfo {author} {\bibfnamefont {D.}~\bibnamefont {Squizzato}},
  \bibinfo {author} {\bibfnamefont {F.}~\bibnamefont {Baboux}}, \bibinfo
  {author} {\bibfnamefont {I.}~\bibnamefont {Amelio}}, \bibinfo {author}
  {\bibfnamefont {A.}~\bibnamefont {Lema{\^\i}tre}}, \bibinfo {author}
  {\bibfnamefont {M.}~\bibnamefont {Morassi}}, \bibinfo {author} {\bibfnamefont
  {I.}~\bibnamefont {Sagnes}}, \bibinfo {author} {\bibfnamefont
  {L.}~\bibnamefont {Le~Gratiet}}, \bibinfo {author} {\bibfnamefont
  {A.}~\bibnamefont {Harouri}}, \bibinfo {author} {\bibfnamefont
  {M.}~\bibnamefont {Wouters}}, \bibinfo {author} {\bibfnamefont
  {I.}~\bibnamefont {Carusotto}}, \bibinfo {author} {\bibfnamefont
  {A.}~\bibnamefont {Amo}}, \bibinfo {author} {\bibfnamefont {M.}~\bibnamefont
  {Richard}}, \bibinfo {author} {\bibfnamefont {A.}~\bibnamefont {Minguzzi}},
  \bibinfo {author} {\bibfnamefont {L.}~\bibnamefont {Canet}}, \bibinfo
  {author} {\bibfnamefont {S.}~\bibnamefont {Ravets}},\ and\ \bibinfo {author}
  {\bibfnamefont {J.}~\bibnamefont {Bloch}},\ }\bibfield  {title} {\bibinfo
  {title} {Kardar--{P}arisi--{Z}hang universality in a one-dimensional
  polariton condensate},\ }\href {https://doi.org/10.1038/s41586-022-05001-8}
  {\bibfield  {journal} {\bibinfo  {journal} {Nature}\ }\textbf {\bibinfo
  {volume} {608}},\ \bibinfo {pages} {687} (\bibinfo {year}
  {2022})}\BibitemShut {NoStop}%
\bibitem [{\citenamefont {Canet}\ \emph {et~al.}(2010)\citenamefont {Canet},
  \citenamefont {Chat\'e}, \citenamefont {Delamotte},\ and\ \citenamefont
  {Wschebor}}]{Canetetal2010}%
  \BibitemOpen
  \bibfield  {author} {\bibinfo {author} {\bibfnamefont {L.}~\bibnamefont
  {Canet}}, \bibinfo {author} {\bibfnamefont {H.}~\bibnamefont {Chat\'e}},
  \bibinfo {author} {\bibfnamefont {B.}~\bibnamefont {Delamotte}},\ and\
  \bibinfo {author} {\bibfnamefont {N.}~\bibnamefont {Wschebor}},\ }\bibfield
  {title} {\bibinfo {title} {Nonperturbative renormalization group for the
  {K}ardar-{P}arisi-{Z}hang equation},\ }\href
  {https://doi.org/10.1103/PhysRevLett.104.150601} {\bibfield  {journal}
  {\bibinfo  {journal} {Phys. Rev. Lett.}\ }\textbf {\bibinfo {volume} {104}},\
  \bibinfo {pages} {150601} (\bibinfo {year} {2010})}\BibitemShut {NoStop}%
\bibitem [{\citenamefont {Kloss}\ \emph {et~al.}(2012)\citenamefont {Kloss},
  \citenamefont {Canet},\ and\ \citenamefont
  {Wschebor}}]{KlossCanetWschebor2012}%
  \BibitemOpen
  \bibfield  {author} {\bibinfo {author} {\bibfnamefont {T.}~\bibnamefont
  {Kloss}}, \bibinfo {author} {\bibfnamefont {L.}~\bibnamefont {Canet}},\ and\
  \bibinfo {author} {\bibfnamefont {N.}~\bibnamefont {Wschebor}},\ }\bibfield
  {title} {\bibinfo {title} {Nonperturbative renormalization group for the
  stationary {K}ardar-{P}arisi-{Z}hang equation: Scaling functions and
  amplitude ratios in $1+1$, $2+1$, and $3+1$ dimensions},\ }\href
  {https://link.aps.org/doi/10.1103/PhysRevE.86.051124} {\bibfield  {journal}
  {\bibinfo  {journal} {Phys. Rev. E}\ }\textbf {\bibinfo {volume} {86}},\
  \bibinfo {pages} {051124} (\bibinfo {year} {2012})}\BibitemShut {NoStop}%
\bibitem [{\citenamefont {Kloss}\ \emph
  {et~al.}(2014{\natexlab{a}})\citenamefont {Kloss}, \citenamefont {Canet},\
  and\ \citenamefont {Wschebor}}]{KlossCanetWschebor2014}%
  \BibitemOpen
  \bibfield  {author} {\bibinfo {author} {\bibfnamefont {T.}~\bibnamefont
  {Kloss}}, \bibinfo {author} {\bibfnamefont {L.}~\bibnamefont {Canet}},\ and\
  \bibinfo {author} {\bibfnamefont {N.}~\bibnamefont {Wschebor}},\ }\bibfield
  {title} {\bibinfo {title} {Strong-coupling phases of the anisotropic
  {K}ardar-{P}arisi-{Z}hang equation},\ }\href
  {https://doi.org/10.1103/PhysRevE.90.062133} {\bibfield  {journal} {\bibinfo
  {journal} {Phys. Rev. E}\ }\textbf {\bibinfo {volume} {90}},\ \bibinfo
  {pages} {062133} (\bibinfo {year} {2014}{\natexlab{a}})}\BibitemShut
  {NoStop}%
\bibitem [{\citenamefont {Kloss}\ \emph
  {et~al.}(2014{\natexlab{b}})\citenamefont {Kloss}, \citenamefont {Canet},
  \citenamefont {Delamotte},\ and\ \citenamefont
  {Wschebor}}]{KlossCanetDelamotteWschebor2014}%
  \BibitemOpen
  \bibfield  {author} {\bibinfo {author} {\bibfnamefont {T.}~\bibnamefont
  {Kloss}}, \bibinfo {author} {\bibfnamefont {L.}~\bibnamefont {Canet}},
  \bibinfo {author} {\bibfnamefont {B.}~\bibnamefont {Delamotte}},\ and\
  \bibinfo {author} {\bibfnamefont {N.}~\bibnamefont {Wschebor}},\ }\bibfield
  {title} {\bibinfo {title} {{K}ardar-{P}arisi-{Z}hang equation with spatially
  correlated noise: A unified picture from nonperturbative {R}enormalization
  {G}roup},\ }\href {https://doi.org/10.1103/PhysRevE.89.022108} {\bibfield
  {journal} {\bibinfo  {journal} {Phys. Rev. E}\ }\textbf {\bibinfo {volume}
  {89}},\ \bibinfo {pages} {022108} (\bibinfo {year}
  {2014}{\natexlab{b}})}\BibitemShut {NoStop}%
\bibitem [{\citenamefont {Zabolitzky}\ and\ \citenamefont
  {Stauffer}(1986)}]{ZabolitzkyStauffer1986}%
  \BibitemOpen
  \bibfield  {author} {\bibinfo {author} {\bibfnamefont {J.~G.}\ \bibnamefont
  {Zabolitzky}}\ and\ \bibinfo {author} {\bibfnamefont {D.}~\bibnamefont
  {Stauffer}},\ }\bibfield  {title} {\bibinfo {title} {Simulation of large
  {E}den clusters},\ }\href {https://doi.org/10.1103/PhysRevA.34.1523}
  {\bibfield  {journal} {\bibinfo  {journal} {Phys. Rev. A}\ }\textbf {\bibinfo
  {volume} {34}},\ \bibinfo {pages} {1523} (\bibinfo {year}
  {1986})}\BibitemShut {NoStop}%
\bibitem [{\citenamefont {Kert{\'e}sz}\ and\ \citenamefont
  {Wolf}(1989)}]{KerteszWolf1989}%
  \BibitemOpen
  \bibfield  {author} {\bibinfo {author} {\bibfnamefont {J.}~\bibnamefont
  {Kert{\'e}sz}}\ and\ \bibinfo {author} {\bibfnamefont {D.~E.}\ \bibnamefont
  {Wolf}},\ }\bibfield  {title} {\bibinfo {title} {Anomalous roughening in
  growth processes},\ }\href {https://doi.org/10.1103/PhysRevLett.62.2571}
  {\bibfield  {journal} {\bibinfo  {journal} {Phys. Rev. Lett.}\ }\textbf
  {\bibinfo {volume} {62}},\ \bibinfo {pages} {2571} (\bibinfo {year}
  {1989})}\BibitemShut {NoStop}%
\bibitem [{\citenamefont {Kim}\ and\ \citenamefont
  {Kosterlitz}(1989)}]{KimKosterlitz1989}%
  \BibitemOpen
  \bibfield  {author} {\bibinfo {author} {\bibfnamefont {J.~M.}\ \bibnamefont
  {Kim}}\ and\ \bibinfo {author} {\bibfnamefont {J.~M.}\ \bibnamefont
  {Kosterlitz}},\ }\bibfield  {title} {\bibinfo {title} {Growth in a restricted
  solid-on-solid model},\ }\href
  {https://link.aps.org/doi/10.1103/PhysRevLett.62.2289} {\bibfield  {journal}
  {\bibinfo  {journal} {Phys. Rev. Lett.}\ }\textbf {\bibinfo {volume} {62}},\
  \bibinfo {pages} {2289} (\bibinfo {year} {1989})}\BibitemShut {NoStop}%
\bibitem [{\citenamefont {Forrest}\ and\ \citenamefont
  {Tang}(1990)}]{ForrestTang1990}%
  \BibitemOpen
  \bibfield  {author} {\bibinfo {author} {\bibfnamefont {B.~M.}\ \bibnamefont
  {Forrest}}\ and\ \bibinfo {author} {\bibfnamefont {L.-H.}\ \bibnamefont
  {Tang}},\ }\bibfield  {title} {\bibinfo {title} {Surface roughening in a
  hypercube-stacking model},\ }\href
  {https://doi.org/10.1103/PhysRevLett.64.1405} {\bibfield  {journal} {\bibinfo
   {journal} {Phys. Rev. Lett.}\ }\textbf {\bibinfo {volume} {64}},\ \bibinfo
  {pages} {1405} (\bibinfo {year} {1990})}\BibitemShut {NoStop}%
\bibitem [{\citenamefont {Kim}\ \emph {et~al.}(1991)\citenamefont {Kim},
  \citenamefont {Kosterlitz},\ and\ \citenamefont {Ala-Nissila}}]{Kim1991}%
  \BibitemOpen
  \bibfield  {author} {\bibinfo {author} {\bibfnamefont {J.~M.}\ \bibnamefont
  {Kim}}, \bibinfo {author} {\bibfnamefont {J.~M.}\ \bibnamefont
  {Kosterlitz}},\ and\ \bibinfo {author} {\bibfnamefont {T.}~\bibnamefont
  {Ala-Nissila}},\ }\bibfield  {title} {\bibinfo {title} {Surface growth and
  crossover behaviour in a restricted solid-on-solid model},\ }\href
  {https://doi.org/10.1088/0305-4470/24/23/022} {\bibfield  {journal} {\bibinfo
   {journal} {Journal of Physics A: Mathematical and General}\ }\textbf
  {\bibinfo {volume} {24}},\ \bibinfo {pages} {5569} (\bibinfo {year}
  {1991})}\BibitemShut {NoStop}%
\bibitem [{\citenamefont {Tang}\ \emph {et~al.}(1992)\citenamefont {Tang},
  \citenamefont {Forrest},\ and\ \citenamefont {Wolf}}]{Tang1992}%
  \BibitemOpen
  \bibfield  {author} {\bibinfo {author} {\bibfnamefont {L.-H.}\ \bibnamefont
  {Tang}}, \bibinfo {author} {\bibfnamefont {B.~M.}\ \bibnamefont {Forrest}},\
  and\ \bibinfo {author} {\bibfnamefont {D.~E.}\ \bibnamefont {Wolf}},\
  }\bibfield  {title} {\bibinfo {title} {Kinetic surface roughening. ii.
  {H}ypercube-stacking models},\ }\href
  {https://doi.org/10.1103/PhysRevA.45.7162} {\bibfield  {journal} {\bibinfo
  {journal} {Phys. Rev. A}\ }\textbf {\bibinfo {volume} {45}},\ \bibinfo
  {pages} {7162} (\bibinfo {year} {1992})}\BibitemShut {NoStop}%
\bibitem [{\citenamefont {Chin}\ and\ \citenamefont {den
  Nijs}(1999)}]{ChinNijs1999}%
  \BibitemOpen
  \bibfield  {author} {\bibinfo {author} {\bibfnamefont {C.-S.}\ \bibnamefont
  {Chin}}\ and\ \bibinfo {author} {\bibfnamefont {M.}~\bibnamefont {den
  Nijs}},\ }\bibfield  {title} {\bibinfo {title} {Stationary-state skewness in
  two-dimensional {K}ardar-{P}arisi-{Z}hang type growth},\ }\href
  {https://doi.org/10.1103/PhysRevE.59.2633} {\bibfield  {journal} {\bibinfo
  {journal} {Phys. Rev. E}\ }\textbf {\bibinfo {volume} {59}},\ \bibinfo
  {pages} {2633} (\bibinfo {year} {1999})}\BibitemShut {NoStop}%
\bibitem [{\citenamefont {Kondev}\ \emph {et~al.}(2000)\citenamefont {Kondev},
  \citenamefont {Henley},\ and\ \citenamefont
  {Salinas}}]{KondevHenleySalinas2000}%
  \BibitemOpen
  \bibfield  {author} {\bibinfo {author} {\bibfnamefont {J.}~\bibnamefont
  {Kondev}}, \bibinfo {author} {\bibfnamefont {C.~L.}\ \bibnamefont {Henley}},\
  and\ \bibinfo {author} {\bibfnamefont {D.~G.}\ \bibnamefont {Salinas}},\
  }\bibfield  {title} {\bibinfo {title} {Nonlinear measures for characterizing
  rough surface morphologies},\ }\href
  {https://doi.org/10.1103/PhysRevE.61.104} {\bibfield  {journal} {\bibinfo
  {journal} {Phys. Rev. E}\ }\textbf {\bibinfo {volume} {61}},\ \bibinfo
  {pages} {104} (\bibinfo {year} {2000})}\BibitemShut {NoStop}%
\bibitem [{\citenamefont {Marinari}\ \emph {et~al.}(2000)\citenamefont
  {Marinari}, \citenamefont {Pagnani},\ and\ \citenamefont
  {Parisi}}]{MarinariPagnaniParisi2000}%
  \BibitemOpen
  \bibfield  {author} {\bibinfo {author} {\bibfnamefont {E.}~\bibnamefont
  {Marinari}}, \bibinfo {author} {\bibfnamefont {A.}~\bibnamefont {Pagnani}},\
  and\ \bibinfo {author} {\bibfnamefont {G.}~\bibnamefont {Parisi}},\
  }\bibfield  {title} {\bibinfo {title} {Critical exponents of the {KPZ}
  equation via multi-surface coding numerical simulations},\ }\href
  {https://doi.org/10.1088/0305-4470/33/46/303} {\bibfield  {journal} {\bibinfo
   {journal} {Journal of Physics A: Mathematical and General}\ }\textbf
  {\bibinfo {volume} {33}},\ \bibinfo {pages} {8181} (\bibinfo {year}
  {2000})}\BibitemShut {NoStop}%
\bibitem [{\citenamefont {Aar{\~a}o~Reis}(2001)}]{AaraoReis2001}%
  \BibitemOpen
  \bibfield  {author} {\bibinfo {author} {\bibfnamefont {F.~D.~A.}\
  \bibnamefont {Aar{\~a}o~Reis}},\ }\bibfield  {title} {\bibinfo {title}
  {Universality and corrections to scaling in the ballistic deposition model},\
  }\href {https://doi.org/10.1103/PhysRevE.63.056116} {\bibfield  {journal}
  {\bibinfo  {journal} {Phys. Rev. E}\ }\textbf {\bibinfo {volume} {63}},\
  \bibinfo {pages} {056116} (\bibinfo {year} {2001})}\BibitemShut {NoStop}%
\bibitem [{\citenamefont {{\'O}dor}\ \emph {et~al.}(2009)\citenamefont
  {{\'O}dor}, \citenamefont {Liedke},\ and\ \citenamefont
  {Heinig}}]{OdorLiedkeHeinig2009}%
  \BibitemOpen
  \bibfield  {author} {\bibinfo {author} {\bibfnamefont {G.}~\bibnamefont
  {{\'O}dor}}, \bibinfo {author} {\bibfnamefont {B.}~\bibnamefont {Liedke}},\
  and\ \bibinfo {author} {\bibfnamefont {K.-H.}\ \bibnamefont {Heinig}},\
  }\bibfield  {title} {\bibinfo {title} {Mapping of (2+1)-dimensional
  {K}ardar-{P}arisi-{Z}hang growth onto a driven lattice gas model of dimers},\
  }\href {https://doi.org/10.1103/PhysRevE.79.021125} {\bibfield  {journal}
  {\bibinfo  {journal} {Phys. Rev. E}\ }\textbf {\bibinfo {volume} {79}},\
  \bibinfo {pages} {021125} (\bibinfo {year} {2009})}\BibitemShut {NoStop}%
\bibitem [{\citenamefont {Kelling}\ and\ \citenamefont
  {\'Odor}(2011)}]{KellingOdor2011}%
  \BibitemOpen
  \bibfield  {author} {\bibinfo {author} {\bibfnamefont {J.}~\bibnamefont
  {Kelling}}\ and\ \bibinfo {author} {\bibfnamefont {G.}~\bibnamefont
  {\'Odor}},\ }\bibfield  {title} {\bibinfo {title} {Extremely large-scale
  simulation of a {K}ardar-{P}arisi-{Z}hang model using graphics cards},\
  }\href {https://link.aps.org/doi/10.1103/PhysRevE.84.061150} {\bibfield
  {journal} {\bibinfo  {journal} {Phys. Rev. E}\ }\textbf {\bibinfo {volume}
  {84}},\ \bibinfo {pages} {061150} (\bibinfo {year} {2011})}\BibitemShut
  {NoStop}%
\bibitem [{\citenamefont {Pagnani}\ and\ \citenamefont
  {Parisi}(2015)}]{PagnaniParisi2015}%
  \BibitemOpen
  \bibfield  {author} {\bibinfo {author} {\bibfnamefont {A.}~\bibnamefont
  {Pagnani}}\ and\ \bibinfo {author} {\bibfnamefont {G.}~\bibnamefont
  {Parisi}},\ }\bibfield  {title} {\bibinfo {title} {Numerical estimate of the
  {K}ardar-{P}arisi-{Z}hang universality class in ($2+1$) dimensions},\ }\href
  {https://link.aps.org/doi/10.1103/PhysRevE.92.010101} {\bibfield  {journal}
  {\bibinfo  {journal} {Phys. Rev. E}\ }\textbf {\bibinfo {volume} {92}},\
  \bibinfo {pages} {010101} (\bibinfo {year} {2015})}\BibitemShut {NoStop}%
\bibitem [{\citenamefont {Miranda}\ and\ \citenamefont
  {Aar{\~a}o~Reis}(2008)}]{MirandaAaraoReis2008}%
  \BibitemOpen
  \bibfield  {author} {\bibinfo {author} {\bibfnamefont {V.~G.}\ \bibnamefont
  {Miranda}}\ and\ \bibinfo {author} {\bibfnamefont {F.~D.~A.}\ \bibnamefont
  {Aar{\~a}o~Reis}},\ }\bibfield  {title} {\bibinfo {title} {Numerical study of
  the {K}ardar-{P}arisi-{Z}hang equation},\ }\href
  {https://doi.org/10.1103/PhysRevE.77.031134} {\bibfield  {journal} {\bibinfo
  {journal} {Phys. Rev. E}\ }\textbf {\bibinfo {volume} {77}},\ \bibinfo
  {pages} {031134} (\bibinfo {year} {2008})}\BibitemShut {NoStop}%
\bibitem [{\citenamefont {Newman}\ and\ \citenamefont
  {Bray}(1996)}]{NewmanBray1996}%
  \BibitemOpen
  \bibfield  {author} {\bibinfo {author} {\bibfnamefont {T.~J.}\ \bibnamefont
  {Newman}}\ and\ \bibinfo {author} {\bibfnamefont {A.~J.}\ \bibnamefont
  {Bray}},\ }\bibfield  {title} {\bibinfo {title} {Strong-coupling behaviour in
  discrete {K}ardar-{P}arisi-{Z}hang equations},\ }\href
  {https://doi.org/10.1088/0305-4470/29/24/016} {\bibfield  {journal} {\bibinfo
   {journal} {Journal of Physics A: Mathematical and General}\ }\textbf
  {\bibinfo {volume} {29}},\ \bibinfo {pages} {7917} (\bibinfo {year}
  {1996})}\BibitemShut {NoStop}%
\bibitem [{\citenamefont {Halpin-Healy}(2012)}]{HalpinHealy2012}%
  \BibitemOpen
  \bibfield  {author} {\bibinfo {author} {\bibfnamefont {T.}~\bibnamefont
  {Halpin-Healy}},\ }\bibfield  {title} {\bibinfo {title} {($2 +
  1$)-dimensional directed polymer in a random medium: {S}caling phenomena and
  universal distributions},\ }\href
  {https://doi.org/10.1103/PhysRevLett.109.170602} {\bibfield  {journal}
  {\bibinfo  {journal} {Phys. Rev. Lett.}\ }\textbf {\bibinfo {volume} {109}},\
  \bibinfo {pages} {170602} (\bibinfo {year} {2012})}\BibitemShut {NoStop}%
\bibitem [{\citenamefont {Oliveira}\ \emph {et~al.}(2013)\citenamefont
  {Oliveira}, \citenamefont {Alves},\ and\ \citenamefont
  {Ferreira}}]{OliveiraAlvesFerreira2013}%
  \BibitemOpen
  \bibfield  {author} {\bibinfo {author} {\bibfnamefont {T.~J.}\ \bibnamefont
  {Oliveira}}, \bibinfo {author} {\bibfnamefont {S.~G.}\ \bibnamefont
  {Alves}},\ and\ \bibinfo {author} {\bibfnamefont {S.~C.}\ \bibnamefont
  {Ferreira}},\ }\bibfield  {title} {\bibinfo {title} {Kardar-{P}arisi-{Z}hang
  universality class in (2+1) dimensions: {U}niversal geometry-dependent
  distributions and finite-time corrections},\ }\href
  {https://doi.org/10.1103/PhysRevE.87.040102} {\bibfield  {journal} {\bibinfo
  {journal} {Phys. Rev. E}\ }\textbf {\bibinfo {volume} {87}},\ \bibinfo
  {pages} {040102} (\bibinfo {year} {2013})}\BibitemShut {NoStop}%
\bibitem [{\citenamefont {Dagvadorj}\ \emph {et~al.}(2015)\citenamefont
  {Dagvadorj}, \citenamefont {Fellows}, \citenamefont
  {Matyja\ifmmode~\acute{s}\else \'{s}\fi{}kiewicz}, \citenamefont {Marchetti},
  \citenamefont {Carusotto},\ and\ \citenamefont {Szyma\ifmmode~\acute{n}\else
  \'{n}\fi{}ska}}]{Dagvadorj2015}%
  \BibitemOpen
  \bibfield  {author} {\bibinfo {author} {\bibfnamefont {G.}~\bibnamefont
  {Dagvadorj}}, \bibinfo {author} {\bibfnamefont {J.~M.}\ \bibnamefont
  {Fellows}}, \bibinfo {author} {\bibfnamefont {S.}~\bibnamefont
  {Matyja\ifmmode~\acute{s}\else \'{s}\fi{}kiewicz}}, \bibinfo {author}
  {\bibfnamefont {F.~M.}\ \bibnamefont {Marchetti}}, \bibinfo {author}
  {\bibfnamefont {I.}~\bibnamefont {Carusotto}},\ and\ \bibinfo {author}
  {\bibfnamefont {M.~H.}\ \bibnamefont {Szyma\ifmmode~\acute{n}\else
  \'{n}\fi{}ska}},\ }\bibfield  {title} {\bibinfo {title} {Nonequilibrium phase
  transition in a two-dimensional driven open quantum system},\ }\href
  {https://doi.org/10.1103/PhysRevX.5.041028} {\bibfield  {journal} {\bibinfo
  {journal} {Phys. Rev. X}\ }\textbf {\bibinfo {volume} {5}},\ \bibinfo {pages}
  {041028} (\bibinfo {year} {2015})}\BibitemShut {NoStop}%
\bibitem [{\citenamefont {Zamora}\ \emph {et~al.}(2017)\citenamefont {Zamora},
  \citenamefont {Sieberer}, \citenamefont {Dunnett}, \citenamefont {Diehl},\
  and\ \citenamefont {Szyma\ifmmode~\acute{n}\else
  \'{n}\fi{}ska}}]{ZamoraSiebererDunnettDiehlSzymanska2017}%
  \BibitemOpen
  \bibfield  {author} {\bibinfo {author} {\bibfnamefont {A.}~\bibnamefont
  {Zamora}}, \bibinfo {author} {\bibfnamefont {L.~M.}\ \bibnamefont
  {Sieberer}}, \bibinfo {author} {\bibfnamefont {K.}~\bibnamefont {Dunnett}},
  \bibinfo {author} {\bibfnamefont {S.}~\bibnamefont {Diehl}},\ and\ \bibinfo
  {author} {\bibfnamefont {M.~H.}\ \bibnamefont {Szyma\ifmmode~\acute{n}\else
  \'{n}\fi{}ska}},\ }\bibfield  {title} {\bibinfo {title} {Tuning across
  universalities with a driven open condensate},\ }\href
  {https://doi.org/10.1103/PhysRevX.7.041006} {\bibfield  {journal} {\bibinfo
  {journal} {Phys. Rev. X}\ }\textbf {\bibinfo {volume} {7}},\ \bibinfo {pages}
  {041006} (\bibinfo {year} {2017})}\BibitemShut {NoStop}%
\bibitem [{\citenamefont {Dunnett}\ \emph {et~al.}(2018)\citenamefont
  {Dunnett}, \citenamefont {Ferrier}, \citenamefont {Zamora}, \citenamefont
  {Dagvadorj},\ and\ \citenamefont {Szyma\ifmmode~\acute{n}\else
  \'{n}\fi{}ska}}]{Dunnett2018}%
  \BibitemOpen
  \bibfield  {author} {\bibinfo {author} {\bibfnamefont {K.}~\bibnamefont
  {Dunnett}}, \bibinfo {author} {\bibfnamefont {A.}~\bibnamefont {Ferrier}},
  \bibinfo {author} {\bibfnamefont {A.}~\bibnamefont {Zamora}}, \bibinfo
  {author} {\bibfnamefont {G.}~\bibnamefont {Dagvadorj}},\ and\ \bibinfo
  {author} {\bibfnamefont {M.~H.}\ \bibnamefont {Szyma\ifmmode~\acute{n}\else
  \'{n}\fi{}ska}},\ }\bibfield  {title} {\bibinfo {title} {Properties of the
  signal mode in the polariton optical parametric oscillator regime},\ }\href
  {https://doi.org/10.1103/PhysRevB.98.165307} {\bibfield  {journal} {\bibinfo
  {journal} {Phys. Rev. B}\ }\textbf {\bibinfo {volume} {98}},\ \bibinfo
  {pages} {165307} (\bibinfo {year} {2018})}\BibitemShut {NoStop}%
\bibitem [{\citenamefont {Ferrier}\ \emph {et~al.}(2022)\citenamefont
  {Ferrier}, \citenamefont {Zamora}, \citenamefont {Dagvadorj},\ and\
  \citenamefont {Szyma\ifmmode~\acute{n}\else
  \'{n}\fi{}ska}}]{FerrierEtAl2020}%
  \BibitemOpen
  \bibfield  {author} {\bibinfo {author} {\bibfnamefont {A.}~\bibnamefont
  {Ferrier}}, \bibinfo {author} {\bibfnamefont {A.}~\bibnamefont {Zamora}},
  \bibinfo {author} {\bibfnamefont {G.}~\bibnamefont {Dagvadorj}},\ and\
  \bibinfo {author} {\bibfnamefont {M.~H.}\ \bibnamefont
  {Szyma\ifmmode~\acute{n}\else \'{n}\fi{}ska}},\ }\bibfield  {title} {\bibinfo
  {title} {Searching for the {K}ardar-{P}arisi-{Z}hang phase in microcavity
  polaritons},\ }\href {https://doi.org/10.1103/PhysRevB.105.205301} {\bibfield
   {journal} {\bibinfo  {journal} {Phys. Rev. B}\ }\textbf {\bibinfo {volume}
  {105}},\ \bibinfo {pages} {205301} (\bibinfo {year} {2022})}\BibitemShut
  {NoStop}%
\bibitem [{\citenamefont {Diessel}\ \emph {et~al.}(2022)\citenamefont
  {Diessel}, \citenamefont {Diehl},\ and\ \citenamefont
  {Chiocchetta}}]{DiesselChiocchetta2022}%
  \BibitemOpen
  \bibfield  {author} {\bibinfo {author} {\bibfnamefont {O.~K.}\ \bibnamefont
  {Diessel}}, \bibinfo {author} {\bibfnamefont {S.}~\bibnamefont {Diehl}},\
  and\ \bibinfo {author} {\bibfnamefont {A.}~\bibnamefont {Chiocchetta}},\
  }\bibfield  {title} {\bibinfo {title} {Emergent {K}ardar-{P}arisi-{Z}hang
  phase in quadratically driven condensates},\ }\href
  {https://doi.org/10.1103/PhysRevLett.128.070401} {\bibfield  {journal}
  {\bibinfo  {journal} {Phys. Rev. Lett.}\ }\textbf {\bibinfo {volume} {128}},\
  \bibinfo {pages} {070401} (\bibinfo {year} {2022})}\BibitemShut {NoStop}%
\bibitem [{\citenamefont {Gladilin}\ and\ \citenamefont
  {Wouters}(2017)}]{GladilinWouters2017}%
  \BibitemOpen
  \bibfield  {author} {\bibinfo {author} {\bibfnamefont {V.~N.}\ \bibnamefont
  {Gladilin}}\ and\ \bibinfo {author} {\bibfnamefont {M.}~\bibnamefont
  {Wouters}},\ }\bibfield  {title} {\bibinfo {title} {Interaction and motion of
  vortices in nonequilibrium quantum fluids},\ }\href
  {https://doi.org/10.1088/1367-2630/aa83a1} {\bibfield  {journal} {\bibinfo
  {journal} {New Journal of Physics}\ }\textbf {\bibinfo {volume} {19}},\
  \bibinfo {pages} {105005} (\bibinfo {year} {2017})}\BibitemShut {NoStop}%
\bibitem [{\citenamefont {Gladilin}\ and\ \citenamefont
  {Wouters}(2019{\natexlab{a}})}]{GladilinWouters2019A}%
  \BibitemOpen
  \bibfield  {author} {\bibinfo {author} {\bibfnamefont {V.~N.}\ \bibnamefont
  {Gladilin}}\ and\ \bibinfo {author} {\bibfnamefont {M.}~\bibnamefont
  {Wouters}},\ }\bibfield  {title} {\bibinfo {title} {Multivortex states and
  dynamics in nonequilibrium polariton condensates},\ }\href
  {https://doi.org/10.1088/1751-8121/ab3abc} {\bibfield  {journal} {\bibinfo
  {journal} {Journal of Physics A: Mathematical and Theoretical}\ }\textbf
  {\bibinfo {volume} {52}},\ \bibinfo {pages} {395303} (\bibinfo {year}
  {2019}{\natexlab{a}})}\BibitemShut {NoStop}%
\bibitem [{\citenamefont {Wachtel}\ \emph {et~al.}(2016)\citenamefont
  {Wachtel}, \citenamefont {Sieberer}, \citenamefont {Diehl},\ and\
  \citenamefont {Altman}}]{WachtelEtAl2016}%
  \BibitemOpen
  \bibfield  {author} {\bibinfo {author} {\bibfnamefont {G.}~\bibnamefont
  {Wachtel}}, \bibinfo {author} {\bibfnamefont {L.~M.}\ \bibnamefont
  {Sieberer}}, \bibinfo {author} {\bibfnamefont {S.}~\bibnamefont {Diehl}},\
  and\ \bibinfo {author} {\bibfnamefont {E.}~\bibnamefont {Altman}},\
  }\bibfield  {title} {\bibinfo {title} {Electrodynamic duality and vortex
  unbinding in driven-dissipative condensates},\ }\href
  {https://doi.org/10.1103/PhysRevB.94.104520} {\bibfield  {journal} {\bibinfo
  {journal} {Phys. Rev. B}\ }\textbf {\bibinfo {volume} {94}},\ \bibinfo
  {pages} {104520} (\bibinfo {year} {2016})}\BibitemShut {NoStop}%
\bibitem [{\citenamefont {Sieberer}\ \emph {et~al.}(2016)\citenamefont
  {Sieberer}, \citenamefont {Wachtel}, \citenamefont {Altman},\ and\
  \citenamefont {Diehl}}]{SiebererWachtelAltmanDiehl2016}%
  \BibitemOpen
  \bibfield  {author} {\bibinfo {author} {\bibfnamefont {L.~M.}\ \bibnamefont
  {Sieberer}}, \bibinfo {author} {\bibfnamefont {G.}~\bibnamefont {Wachtel}},
  \bibinfo {author} {\bibfnamefont {E.}~\bibnamefont {Altman}},\ and\ \bibinfo
  {author} {\bibfnamefont {S.}~\bibnamefont {Diehl}},\ }\bibfield  {title}
  {\bibinfo {title} {Lattice duality for the compact {K}ardar-{P}arisi-{Z}hang
  equation},\ }\href {https://doi.org/10.1103/PhysRevB.94.104521} {\bibfield
  {journal} {\bibinfo  {journal} {Phys. Rev. B}\ }\textbf {\bibinfo {volume}
  {94}},\ \bibinfo {pages} {104521} (\bibinfo {year} {2016})}\BibitemShut
  {NoStop}%
\bibitem [{\citenamefont {Szyma{\'n}ska}\ \emph {et~al.}(2006)\citenamefont
  {Szyma{\'n}ska}, \citenamefont {Keeling},\ and\ \citenamefont
  {Littlewood}}]{Szymanska2006}%
  \BibitemOpen
  \bibfield  {author} {\bibinfo {author} {\bibfnamefont {M.~H.}\ \bibnamefont
  {Szyma{\'n}ska}}, \bibinfo {author} {\bibfnamefont {J.}~\bibnamefont
  {Keeling}},\ and\ \bibinfo {author} {\bibfnamefont {P.~B.}\ \bibnamefont
  {Littlewood}},\ }\bibfield  {title} {\bibinfo {title} {Nonequilibrium quantum
  condensation in an incoherently pumped dissipative system},\ }\href
  {https://doi.org/10.1103/PhysRevLett.96.230602} {\bibfield  {journal}
  {\bibinfo  {journal} {Phys. Rev. Lett.}\ }\textbf {\bibinfo {volume} {96}},\
  \bibinfo {pages} {230602} (\bibinfo {year} {2006})}\BibitemShut {NoStop}%
\bibitem [{\citenamefont {Caputo}\ \emph {et~al.}(2018)\citenamefont {Caputo},
  \citenamefont {Ballarini}, \citenamefont {Dagvadorj}, \citenamefont
  {S{\'a}nchez~Mu{\~n}oz}, \citenamefont {De~Giorgi}, \citenamefont {Dominici},
  \citenamefont {West}, \citenamefont {Pfeiffer}, \citenamefont {Gigli},
  \citenamefont {Laussy}, \citenamefont {Szyma{\'n}ska},\ and\ \citenamefont
  {Sanvitto}}]{Caputo2017}%
  \BibitemOpen
  \bibfield  {author} {\bibinfo {author} {\bibfnamefont {D.}~\bibnamefont
  {Caputo}}, \bibinfo {author} {\bibfnamefont {D.}~\bibnamefont {Ballarini}},
  \bibinfo {author} {\bibfnamefont {G.}~\bibnamefont {Dagvadorj}}, \bibinfo
  {author} {\bibfnamefont {C.}~\bibnamefont {S{\'a}nchez~Mu{\~n}oz}}, \bibinfo
  {author} {\bibfnamefont {M.}~\bibnamefont {De~Giorgi}}, \bibinfo {author}
  {\bibfnamefont {L.}~\bibnamefont {Dominici}}, \bibinfo {author}
  {\bibfnamefont {K.}~\bibnamefont {West}}, \bibinfo {author} {\bibfnamefont
  {L.~N.}\ \bibnamefont {Pfeiffer}}, \bibinfo {author} {\bibfnamefont
  {G.}~\bibnamefont {Gigli}}, \bibinfo {author} {\bibfnamefont {F.~P.}\
  \bibnamefont {Laussy}}, \bibinfo {author} {\bibfnamefont {M.~H.}\
  \bibnamefont {Szyma{\'n}ska}},\ and\ \bibinfo {author} {\bibfnamefont
  {D.}~\bibnamefont {Sanvitto}},\ }\bibfield  {title} {\bibinfo {title}
  {Topological order and thermal equilibrium in polariton condensates},\ }\href
  {https://doi.org/10.1038/nmat5039} {\bibfield  {journal} {\bibinfo  {journal}
  {Nature Materials}\ }\textbf {\bibinfo {volume} {17}},\ \bibinfo {pages}
  {145} (\bibinfo {year} {2018})}\BibitemShut {NoStop}%
\bibitem [{\citenamefont {Gladilin}\ and\ \citenamefont
  {Wouters}(2019{\natexlab{b}})}]{GladilinWouters2019B}%
  \BibitemOpen
  \bibfield  {author} {\bibinfo {author} {\bibfnamefont {V.~N.}\ \bibnamefont
  {Gladilin}}\ and\ \bibinfo {author} {\bibfnamefont {M.}~\bibnamefont
  {Wouters}},\ }\bibfield  {title} {\bibinfo {title} {Noise-induced transition
  from superfluid to vortex state in two-dimensional nonequilibrium polariton
  condensates},\ }\href {https://doi.org/10.1103/PhysRevB.100.214506}
  {\bibfield  {journal} {\bibinfo  {journal} {Phys. Rev. B}\ }\textbf {\bibinfo
  {volume} {100}},\ \bibinfo {pages} {214506} (\bibinfo {year}
  {2019}{\natexlab{b}})}\BibitemShut {NoStop}%
\bibitem [{\citenamefont {Comaron}\ \emph {et~al.}(2021)\citenamefont
  {Comaron}, \citenamefont {Carusotto}, \citenamefont {Szyma{\'n}ska},\ and\
  \citenamefont {Proukakis}}]{Comaron2021}%
  \BibitemOpen
  \bibfield  {author} {\bibinfo {author} {\bibfnamefont {P.}~\bibnamefont
  {Comaron}}, \bibinfo {author} {\bibfnamefont {I.}~\bibnamefont {Carusotto}},
  \bibinfo {author} {\bibfnamefont {M.~H.}\ \bibnamefont {Szyma{\'n}ska}},\
  and\ \bibinfo {author} {\bibfnamefont {N.~P.}\ \bibnamefont {Proukakis}},\
  }\bibfield  {title} {\bibinfo {title} {Non-equilibrium
  {B}erezinskii-{K}osterlitz-{T}houless transition in driven-dissipative
  condensates},\ }\href {https://doi.org/10.1209/0295-5075/133/17002}
  {\bibfield  {journal} {\bibinfo  {journal} {EPL (Europhysics Letters)}\
  }\textbf {\bibinfo {volume} {133}},\ \bibinfo {pages} {17002} (\bibinfo
  {year} {2021})}\BibitemShut {NoStop}%
\bibitem [{\citenamefont {Gladilin}\ and\ \citenamefont
  {Wouters}(2021)}]{GladilinWouters2021}%
  \BibitemOpen
  \bibfield  {author} {\bibinfo {author} {\bibfnamefont {V.~N.}\ \bibnamefont
  {Gladilin}}\ and\ \bibinfo {author} {\bibfnamefont {M.}~\bibnamefont
  {Wouters}},\ }\bibfield  {title} {\bibinfo {title} {Vortex unbinding
  transition in nonequilibrium photon condensates},\ }\href
  {https://doi.org/10.1103/PhysRevA.104.043516} {\bibfield  {journal} {\bibinfo
   {journal} {Phys. Rev. A}\ }\textbf {\bibinfo {volume} {104}},\ \bibinfo
  {pages} {043516} (\bibinfo {year} {2021})}\BibitemShut {NoStop}%
\bibitem [{\citenamefont {Mei}\ \emph {et~al.}(2021)\citenamefont {Mei},
  \citenamefont {Ji},\ and\ \citenamefont {Wouters}}]{MeiJiWouters2021}%
  \BibitemOpen
  \bibfield  {author} {\bibinfo {author} {\bibfnamefont {Q.}~\bibnamefont
  {Mei}}, \bibinfo {author} {\bibfnamefont {K.}~\bibnamefont {Ji}},\ and\
  \bibinfo {author} {\bibfnamefont {M.}~\bibnamefont {Wouters}},\ }\bibfield
  {title} {\bibinfo {title} {Spatiotemporal scaling of two-dimensional
  nonequilibrium exciton-polariton systems with weak interactions},\ }\href
  {https://doi.org/10.1103/PhysRevB.103.045302} {\bibfield  {journal} {\bibinfo
   {journal} {Phys. Rev. B}\ }\textbf {\bibinfo {volume} {103}},\ \bibinfo
  {pages} {045302} (\bibinfo {year} {2021})}\BibitemShut {NoStop}%
\bibitem [{\citenamefont {Schneider}\ \emph {et~al.}(2016)\citenamefont
  {Schneider}, \citenamefont {Winkler}, \citenamefont {Fraser}, \citenamefont
  {Kamp}, \citenamefont {Yamamoto}, \citenamefont {Ostrovskaya},\ and\
  \citenamefont {Höfling}}]{Schneider_2017}%
  \BibitemOpen
  \bibfield  {author} {\bibinfo {author} {\bibfnamefont {C.}~\bibnamefont
  {Schneider}}, \bibinfo {author} {\bibfnamefont {K.}~\bibnamefont {Winkler}},
  \bibinfo {author} {\bibfnamefont {M.~D.}\ \bibnamefont {Fraser}}, \bibinfo
  {author} {\bibfnamefont {M.}~\bibnamefont {Kamp}}, \bibinfo {author}
  {\bibfnamefont {Y.}~\bibnamefont {Yamamoto}}, \bibinfo {author}
  {\bibfnamefont {E.~A.}\ \bibnamefont {Ostrovskaya}},\ and\ \bibinfo {author}
  {\bibfnamefont {S.}~\bibnamefont {Höfling}},\ }\bibfield  {title} {\bibinfo
  {title} {Exciton-polariton trapping and potential landscape engineering},\
  }\href {https://doi.org/10.1088/0034-4885/80/1/016503} {\bibfield  {journal}
  {\bibinfo  {journal} {Reports on Progress in Physics}\ }\textbf {\bibinfo
  {volume} {80}},\ \bibinfo {pages} {016503} (\bibinfo {year}
  {2016})}\BibitemShut {NoStop}%
\bibitem [{\citenamefont {Halpin-Healy}(2013)}]{HalpinHealy2013}%
  \BibitemOpen
  \bibfield  {author} {\bibinfo {author} {\bibfnamefont {T.}~\bibnamefont
  {Halpin-Healy}},\ }\bibfield  {title} {\bibinfo {title} {Extremal paths, the
  stochastic heat equation, and the three-dimensional {K}ardar-{P}arisi-{Z}hang
  universality class},\ }\href {https://doi.org/10.1103/PhysRevE.88.042118}
  {\bibfield  {journal} {\bibinfo  {journal} {Phys. Rev. E}\ }\textbf {\bibinfo
  {volume} {88}},\ \bibinfo {pages} {042118} (\bibinfo {year}
  {2013})}\BibitemShut {NoStop}%
\bibitem [{\citenamefont {Wouters}\ and\ \citenamefont
  {Carusotto}(2007)}]{WoutersCarusotto2007b}%
  \BibitemOpen
  \bibfield  {author} {\bibinfo {author} {\bibfnamefont {M.}~\bibnamefont
  {Wouters}}\ and\ \bibinfo {author} {\bibfnamefont {I.}~\bibnamefont
  {Carusotto}},\ }\bibfield  {title} {\bibinfo {title} {Excitations in a
  nonequilibrium bose-einstein condensate of exciton polaritons},\ }\href
  {https://link.aps.org/doi/10.1103/PhysRevLett.99.140402} {\bibfield
  {journal} {\bibinfo  {journal} {Phys. Rev. Lett.}\ }\textbf {\bibinfo
  {volume} {99}},\ \bibinfo {pages} {140402} (\bibinfo {year}
  {2007})}\BibitemShut {NoStop}%
\bibitem [{\citenamefont {Wouters}\ and\ \citenamefont
  {Savona}(2009)}]{WoutersSavona2009}%
  \BibitemOpen
  \bibfield  {author} {\bibinfo {author} {\bibfnamefont {M.}~\bibnamefont
  {Wouters}}\ and\ \bibinfo {author} {\bibfnamefont {V.}~\bibnamefont
  {Savona}},\ }\bibfield  {title} {\bibinfo {title} {Stochastic classical field
  model for polariton condensates},\ }\href
  {https://doi.org/10.1103/PhysRevB.79.165302} {\bibfield  {journal} {\bibinfo
  {journal} {Phys. Rev. B}\ }\textbf {\bibinfo {volume} {79}},\ \bibinfo
  {pages} {165302} (\bibinfo {year} {2009})}\BibitemShut {NoStop}%
\bibitem [{\citenamefont {Baboux}\ \emph {et~al.}(2018)\citenamefont {Baboux},
  \citenamefont {De~Bernardis}, \citenamefont {Goblot}, \citenamefont
  {Gladilin}, \citenamefont {Gomez}, \citenamefont {Galopin}, \citenamefont
  {Le~Gratiet}, \citenamefont {Lema\^{i}tre}, \citenamefont {Sagnes},
  \citenamefont {Carusotto},\ and\ \citenamefont {et~al.}}]{Baboux2018}%
  \BibitemOpen
  \bibfield  {author} {\bibinfo {author} {\bibfnamefont {F.}~\bibnamefont
  {Baboux}}, \bibinfo {author} {\bibfnamefont {D.}~\bibnamefont
  {De~Bernardis}}, \bibinfo {author} {\bibfnamefont {V.}~\bibnamefont
  {Goblot}}, \bibinfo {author} {\bibfnamefont {V.~N.}\ \bibnamefont
  {Gladilin}}, \bibinfo {author} {\bibfnamefont {C.}~\bibnamefont {Gomez}},
  \bibinfo {author} {\bibfnamefont {E.}~\bibnamefont {Galopin}}, \bibinfo
  {author} {\bibfnamefont {L.}~\bibnamefont {Le~Gratiet}}, \bibinfo {author}
  {\bibfnamefont {A.}~\bibnamefont {Lema\^{i}tre}}, \bibinfo {author}
  {\bibfnamefont {I.}~\bibnamefont {Sagnes}}, \bibinfo {author} {\bibfnamefont
  {I.}~\bibnamefont {Carusotto}},\ and\ \bibinfo {author} {\bibnamefont
  {et~al.}},\ }\bibfield  {title} {\bibinfo {title} {Unstable and stable
  regimes of polariton condensation},\ }\href
  {https://doi.org/10.1364/OPTICA.5.001163} {\bibfield  {journal} {\bibinfo
  {journal} {Optica}\ }\textbf {\bibinfo {volume} {5}},\ \bibinfo {pages}
  {1163} (\bibinfo {year} {2018})}\BibitemShut {NoStop}%
\bibitem [{\citenamefont {Wouters}\ and\ \citenamefont
  {Carusotto}(2010)}]{WoutersCarusotto2010}%
  \BibitemOpen
  \bibfield  {author} {\bibinfo {author} {\bibfnamefont {M.}~\bibnamefont
  {Wouters}}\ and\ \bibinfo {author} {\bibfnamefont {I.}~\bibnamefont
  {Carusotto}},\ }\bibfield  {title} {\bibinfo {title} {Superfluidity and
  critical velocities in nonequilibrium bose-einstein condensates},\ }\href
  {https://link.aps.org/doi/10.1103/PhysRevLett.105.020602} {\bibfield
  {journal} {\bibinfo  {journal} {Phys. Rev. Lett.}\ }\textbf {\bibinfo
  {volume} {105}},\ \bibinfo {pages} {020602} (\bibinfo {year}
  {2010})}\BibitemShut {NoStop}%
\bibitem [{\citenamefont {Chiocchetta}\ and\ \citenamefont
  {Carusotto}(2013)}]{ChiochettaCarusotto2013}%
  \BibitemOpen
  \bibfield  {author} {\bibinfo {author} {\bibfnamefont {A.}~\bibnamefont
  {Chiocchetta}}\ and\ \bibinfo {author} {\bibfnamefont {I.}~\bibnamefont
  {Carusotto}},\ }\bibfield  {title} {\bibinfo {title} {Non-equilibrium
  quasi-condensates in reduced dimensions},\ }\href
  {https://doi.org/10.1209/0295-5075/102/67007} {\bibfield  {journal} {\bibinfo
   {journal} {EPL (Europhysics Letters)}\ }\textbf {\bibinfo {volume} {102}},\
  \bibinfo {pages} {67007} (\bibinfo {year} {2013})}\BibitemShut {NoStop}%
\bibitem [{\citenamefont {Aranson}\ \emph {et~al.}(1998)\citenamefont
  {Aranson}, \citenamefont {Scheidl},\ and\ \citenamefont
  {Vinokur}}]{Aranson1998}%
  \BibitemOpen
  \bibfield  {author} {\bibinfo {author} {\bibfnamefont {I.~S.}\ \bibnamefont
  {Aranson}}, \bibinfo {author} {\bibfnamefont {S.}~\bibnamefont {Scheidl}},\
  and\ \bibinfo {author} {\bibfnamefont {V.~M.}\ \bibnamefont {Vinokur}},\
  }\bibfield  {title} {\bibinfo {title} {Nonequilibrium dislocation dynamics
  and instability of driven vortex lattices in two dimensions},\ }\href
  {https://doi.org/10.1103/PhysRevB.58.14541} {\bibfield  {journal} {\bibinfo
  {journal} {Phys. Rev. B}\ }\textbf {\bibinfo {volume} {58}},\ \bibinfo
  {pages} {14541} (\bibinfo {year} {1998})}\BibitemShut {NoStop}%
\bibitem [{\citenamefont {Chen}\ and\ \citenamefont {Toner}(2013)}]{Chen2013}%
  \BibitemOpen
  \bibfield  {author} {\bibinfo {author} {\bibfnamefont {L.}~\bibnamefont
  {Chen}}\ and\ \bibinfo {author} {\bibfnamefont {J.}~\bibnamefont {Toner}},\
  }\bibfield  {title} {\bibinfo {title} {Universality for moving stripes: {A}
  hydrodynamic theory of polar active smectics},\ }\href
  {https://doi.org/10.1103/PhysRevLett.111.088701} {\bibfield  {journal}
  {\bibinfo  {journal} {Phys. Rev. Lett.}\ }\textbf {\bibinfo {volume} {111}},\
  \bibinfo {pages} {088701} (\bibinfo {year} {2013})}\BibitemShut {NoStop}%
\bibitem [{\citenamefont {Wolf}(1991)}]{Wolf1991}%
  \BibitemOpen
  \bibfield  {author} {\bibinfo {author} {\bibfnamefont {D.~E.}\ \bibnamefont
  {Wolf}},\ }\bibfield  {title} {\bibinfo {title} {Kinetic roughening of
  vicinal surfaces},\ }\href {https://doi.org/10.1103/PhysRevLett.67.1783}
  {\bibfield  {journal} {\bibinfo  {journal} {Phys. Rev. Lett.}\ }\textbf
  {\bibinfo {volume} {67}},\ \bibinfo {pages} {1783} (\bibinfo {year}
  {1991})}\BibitemShut {NoStop}%
\bibitem [{\citenamefont {T\"auber}\ and\ \citenamefont
  {Frey}(2002)}]{TauberFrey2002}%
  \BibitemOpen
  \bibfield  {author} {\bibinfo {author} {\bibfnamefont {U.~C.}\ \bibnamefont
  {T\"auber}}\ and\ \bibinfo {author} {\bibfnamefont {E.}~\bibnamefont
  {Frey}},\ }\bibfield  {title} {\bibinfo {title} {Universality classes in the
  anisotropic {K}ardar-{P}arisi-{Z}hang model},\ }\href
  {http://stacks.iop.org/0295-5075/59/i=5/a=655} {\bibfield  {journal}
  {\bibinfo  {journal} {Europhys. Lett.}\ }\textbf {\bibinfo {volume} {59}},\
  \bibinfo {pages} {655} (\bibinfo {year} {2002})}\BibitemShut {NoStop}%
\bibitem [{\citenamefont {Amar}\ and\ \citenamefont
  {Family}(1990)}]{AmarFamily1990}%
  \BibitemOpen
  \bibfield  {author} {\bibinfo {author} {\bibfnamefont {J.~G.}\ \bibnamefont
  {Amar}}\ and\ \bibinfo {author} {\bibfnamefont {F.}~\bibnamefont {Family}},\
  }\bibfield  {title} {\bibinfo {title} {Numerical solution of a continuum
  equation for interface growth in 2+1 dimensions},\ }\href
  {https://doi.org/10.1103/PhysRevA.41.3399} {\bibfield  {journal} {\bibinfo
  {journal} {Phys. Rev. A}\ }\textbf {\bibinfo {volume} {41}},\ \bibinfo
  {pages} {3399} (\bibinfo {year} {1990})}\BibitemShut {NoStop}%
\bibitem [{\citenamefont {Grossmann}\ \emph {et~al.}(1991)\citenamefont
  {Grossmann}, \citenamefont {Guo},\ and\ \citenamefont
  {Grant}}]{GrossmanGuoGrant1991}%
  \BibitemOpen
  \bibfield  {author} {\bibinfo {author} {\bibfnamefont {B.}~\bibnamefont
  {Grossmann}}, \bibinfo {author} {\bibfnamefont {H.}~\bibnamefont {Guo}},\
  and\ \bibinfo {author} {\bibfnamefont {M.}~\bibnamefont {Grant}},\ }\bibfield
   {title} {\bibinfo {title} {Kinetic roughening of interfaces in driven
  systems},\ }\href {https://doi.org/10.1103/PhysRevA.43.1727} {\bibfield
  {journal} {\bibinfo  {journal} {Phys. Rev. A}\ }\textbf {\bibinfo {volume}
  {43}},\ \bibinfo {pages} {1727} (\bibinfo {year} {1991})}\BibitemShut
  {NoStop}%
\bibitem [{\citenamefont {Moser}\ \emph {et~al.}(1991)\citenamefont {Moser},
  \citenamefont {Kert{\'e}sz},\ and\ \citenamefont
  {Wolf}}]{MoserKerteszWolf1991}%
  \BibitemOpen
  \bibfield  {author} {\bibinfo {author} {\bibfnamefont {K.}~\bibnamefont
  {Moser}}, \bibinfo {author} {\bibfnamefont {J.}~\bibnamefont {Kert{\'e}sz}},\
  and\ \bibinfo {author} {\bibfnamefont {D.~E.}\ \bibnamefont {Wolf}},\
  }\bibfield  {title} {\bibinfo {title} {Numerical solution of the
  {K}ardar-{P}arisi-{Z}hang equation in one, two and three dimensions},\ }\href
  {https://doi.org/https://doi.org/10.1016/0378-4371(91)90017-7} {\bibfield
  {journal} {\bibinfo  {journal} {Physica A: Statistical Mechanics and its
  Applications}\ }\textbf {\bibinfo {volume} {178}},\ \bibinfo {pages} {215}
  (\bibinfo {year} {1991})}\BibitemShut {NoStop}%
\bibitem [{\citenamefont {Wiese}(1998)}]{Wiese1998}%
  \BibitemOpen
  \bibfield  {author} {\bibinfo {author} {\bibfnamefont {K.~J.}\ \bibnamefont
  {Wiese}},\ }\bibfield  {title} {\bibinfo {title} {On the {P}erturbation
  {E}xpansion of the {KPZ} {E}quation},\ }\href
  {https://doi.org/10.1023/B:JOSS.0000026730.76868.c4} {\bibfield  {journal}
  {\bibinfo  {journal} {Journal of Statistical Physics}\ }\textbf {\bibinfo
  {volume} {93}},\ \bibinfo {pages} {143} (\bibinfo {year} {1998})}\BibitemShut
  {NoStop}%
\bibitem [{\citenamefont {Canet}\ \emph {et~al.}(2011)\citenamefont {Canet},
  \citenamefont {Chat\'e}, \citenamefont {Delamotte},\ and\ \citenamefont
  {Wschebor}}]{Canetetal2011}%
  \BibitemOpen
  \bibfield  {author} {\bibinfo {author} {\bibfnamefont {L.}~\bibnamefont
  {Canet}}, \bibinfo {author} {\bibfnamefont {H.}~\bibnamefont {Chat\'e}},
  \bibinfo {author} {\bibfnamefont {B.}~\bibnamefont {Delamotte}},\ and\
  \bibinfo {author} {\bibfnamefont {N.}~\bibnamefont {Wschebor}},\ }\bibfield
  {title} {\bibinfo {title} {Nonperturbative {R}enormalization {G}roup for the
  {K}ardar-{P}arisi-{Z}hang equation: General framework and first
  applications},\ }\href {https://doi.org/10.1103/PhysRevE.84.061128}
  {\bibfield  {journal} {\bibinfo  {journal} {Phys. Rev. E}\ }\textbf {\bibinfo
  {volume} {84}},\ \bibinfo {pages} {061128} (\bibinfo {year}
  {2011})}\BibitemShut {NoStop}%
\bibitem [{\citenamefont {Pr\"ahofer}\ and\ \citenamefont
  {Spohn}(2000)}]{PrahoferSpohn2000}%
  \BibitemOpen
  \bibfield  {author} {\bibinfo {author} {\bibfnamefont {M.}~\bibnamefont
  {Pr\"ahofer}}\ and\ \bibinfo {author} {\bibfnamefont {H.}~\bibnamefont
  {Spohn}},\ }\bibfield  {title} {\bibinfo {title} {Universal distributions for
  growth processes in $1+1$ dimensions and random matrices},\ }\href
  {https://doi.org/10.1103/PhysRevLett.84.4882} {\bibfield  {journal} {\bibinfo
   {journal} {Phys. Rev. Lett}\ }\textbf {\bibinfo {volume} {84}},\ \bibinfo
  {pages} {4882} (\bibinfo {year} {2000})}\BibitemShut {NoStop}%
\bibitem [{Note1()}]{Note1}%
  \BibitemOpen
  \bibinfo {note} {In Ref. \cite {OliveiraAlvesFerreira2013}, the numerical
  data for the distribution of the fluctuations of the 2D KPZ interface is
  fitted using a Gumbel distribution with given parameters. Thus, in Fig.~\ref
  {fig: histogram} we used the same Gumbel function, which is equivalent to
  comparing with the KPZ numerical data on the range probed in our work.
  However, let us note that the Gumbel distribution cannot be expected to be
  the true distribution for the 2D KPZ, and it should not be used to model the
  far tails. Since these tails are not resolved in our data, the comparison
  with the Gumbel function is appropriate.}\BibitemShut {Stop}%
\bibitem [{\citenamefont {Larson}\ and\ \citenamefont
  {Edwards}(2013)}]{LarsonEdwards}%
  \BibitemOpen
  \bibfield  {author} {\bibinfo {author} {\bibfnamefont {R.}~\bibnamefont
  {Larson}}\ and\ \bibinfo {author} {\bibfnamefont {B.~H.}\ \bibnamefont
  {Edwards}},\ }\href@noop {} {\emph {\bibinfo {title} {Multivariate
  calculus}}}\ (\bibinfo  {publisher} {Cengage Learning},\ \bibinfo {year}
  {2013})\BibitemShut {NoStop}%
\bibitem [{\citenamefont {Deligiannis}(2022)}]{konstantinos_thesis}%
  \BibitemOpen
  \bibfield  {author} {\bibinfo {author} {\bibfnamefont {K.}~\bibnamefont
  {Deligiannis}},\ }\bibfield  {title} {\bibinfo {title}
  {{K}ardar-{P}arisi-{Z}hang universality in the phase of a condensate of
  exciton polaritons: from the scaling of correlation functions to advanced
  statistics of the phase in 1+1 and 2+1 dimensions}\ }\href
  {https://doi.org/https://tel.archives-ouvertes.fr/tel-03722989}
  {https://tel.archives-ouvertes.fr/tel-03722989} (\bibinfo {year} {2022}),\
  \bibinfo {note} {universit{\'e} Grenoble - Alpes}\BibitemShut {NoStop}%
\end{thebibliography}

%

\end{document}